%% file: ms.tex
\documentclass{emulateapj}

\def\hda{\object[HD 108317]{HD~108317}}
\def\hdb{\object[HD 128279]{HD~128279}}
\def\bd{\object[BD+17 3248]{BD$+$17~3248}}
\def\hdc{\object[HD 160617]{HD~160617}}
\def\cs{\object[BPS CS 22892-052]{CS~22892--052}}

\def\kmsec{\mbox{km~s$^{\rm -1}$}}
\def\logg{\mbox{log~{\it g}}}
\def\msun{\mbox{$M_{\odot}$}}
\def\teff{\mbox{$T_{\rm eff}$}}
\def\vt{\mbox{$v_{\rm t}$}}
\def\rpro{\mbox{$r$-process}}
\def\spro{\mbox{$s$-process}}

\def\loggf{$\log gf$}
\def\first{1$^{\rm st}$}
\def\second{2$^{\rm nd}$}
\def\third{3$^{\rm rd}$}

\shorttitle{New Arsenic and Selenium Abundances}
\shortauthors{Roederer et al.}

\begin{document}

\title{
New Detections of Arsenic, Selenium, and Other Heavy Elements \\
in Two Metal-Poor Stars\footnotemark[1]}

\footnotetext[1]{
Based on observations made with the NASA/ESA Hubble Space Telescope, 
obtained at the Space Telescope Science Institute, 
which is operated by the Association of Universities for Research in 
Astronomy, Inc., under NASA contract NAS~5-26555. 
These observations are associated with programs GO-12268 and GO-12976.
}

\author{
Ian U.\ Roederer,\altaffilmark{2}
Hendrik Schatz,\altaffilmark{3,}\altaffilmark{4,}\altaffilmark{5}
James E.\ Lawler,\altaffilmark{6}
Timothy C.\ Beers,\altaffilmark{3,}\altaffilmark{4,}\altaffilmark{7}
John J.\ Cowan,\altaffilmark{8} \\
Anna Frebel,\altaffilmark{9}
Inese I.\ Ivans,\altaffilmark{10}
Christopher Sneden,\altaffilmark{11}
Jennifer S.\ Sobeck\altaffilmark{12}
}

\altaffiltext{2}{Department of Astronomy, University of Michigan,
Ann Arbor, MI 48109, USA
}
\altaffiltext{3}{Department of Physics \& Astronomy,
Michigan State University, E.\ Lansing, MI 48824, USA
}
\altaffiltext{4}{Joint Institute for Nuclear Astrophysics, 
Michigan State University, E.\ Lansing, MI  48824, USA
}
\altaffiltext{5}{National Superconducting Cyclotron Laboratory, 
Michigan State University, East Lansing, MI 48824, USA
}
\altaffiltext{6}{Department of Physics, University of Wisconsin, 
Madison, WI 53706, USA
}
\altaffiltext{7}{National Optical Astronomy Observatory, Tucson, AZ 85719,
USA
}
\altaffiltext{8}{Homer L.\ Dodge Department of Physics and Astronomy,
University of Oklahoma, 
Norman, OK 73019, USA
}
\altaffiltext{9}{Department of Physics \& 
Kavli Institute for Astrophysics and Space Research, 
Massachusetts Institute of Technology, 
Cambridge, MA 02139, USA
}
\altaffiltext{10}{Department of Physics and Astronomy, University of Utah,
Salt Lake City, UT 84112, USA
}
\altaffiltext{11}{Department of Astronomy, University of Texas at Austin,
Austin, TX 78712, USA
}
\altaffiltext{12}{Department of Astronomy, University of Virginia,
Charlottesville, VA 22904
}


\addtocounter{footnote}{12}

\begin{abstract}

We use the Space Telescope Imaging Spectrograph 
on board the \textit{Hubble Space Telescope}
to obtain new high-quality spectra covering the
1900~$\leq \lambda \leq$~2360~\AA\
wavelength range for two metal-poor stars,
\mbox{HD~108317} and \mbox{HD~128279}.
We derive abundances of
Cu~\textsc{ii}, Zn~\textsc{ii}, As~\textsc{i}, Se~\textsc{i},
Mo~\textsc{ii}, and Cd~\textsc{ii},
which have not been detected previously in either star.
Abundances derived for
Ge~\textsc{i}, Te~\textsc{i},
Os~\textsc{ii}, and Pt~\textsc{i}
confirm those derived from lines at longer wavelengths.
We also derive upper limits from the non-detection of
W~\textsc{ii}, Hg~\textsc{ii}, Pb~\textsc{ii},
and Bi~\textsc{i}.
The mean [As/Fe] ratio derived from these two stars
and five others in the literature is unchanged
over the metallicity range $-$2.8~$<$~[Fe/H]~$< -$0.6,
$\langle$[As/Fe]$\rangle = +$0.28~$\pm$~0.14 ($\sigma =$~0.36~dex).
The mean [Se/Fe] ratio derived from these two stars
and six others in the literature is also constant,
$\langle$[Se/Fe]$\rangle = +$0.16~$\pm$~0.09 ($\sigma =$~0.26~dex).
The As and Se abundances are enhanced relative to a 
simple extrapolation of the iron-peak abundances to higher masses,
suggesting that this mass region
(75~$\leq A \leq$~82) may be the point at which
a different nucleosynthetic mechanism
begins to dominate the quasi-equilibrium
$\alpha$-rich freezeout of the iron peak.
$\langle$[Cu~\textsc{ii}/Cu~\textsc{i}]$\rangle = +$0.56~$\pm$~0.23
in \hda\ and \hdb, and we infer that
lines of Cu~\textsc{i} may not be formed in 
local thermodynamic equilibrium in these stars.
The [Zn/Fe], [Mo/Fe], [Cd/Fe], and [Os/Fe]
ratios are also derived from neutral and ionized species,
and each ratio pair agrees within the mutual uncertainties,
which range from 0.15 to 0.52~dex.

\end{abstract}

\keywords{
nuclear reactions, nucleosynthesis, abundances ---
stars: abundances ---
stars: individual (HD~108317, HD~128279) ---
stars: Population II
}

\section{Introduction}
\label{intro}

The near ultra-violet (NUV) 
portion of the spectra of late-type (FGK) stars
contains absorption lines of
atomic species that are not accessible
in the optical or near-infrared domains.
In the last two decades, 
NUV data obtained with 
the Goddard High Resolution Spectrograph
and Space Telescope Imaging Spectrograph (STIS)
on board the 
\textit{Hubble Space Telescope} (\textit{HST})
have enabled the first compelling detections of
numerous elements heavier than the iron-group
in late-type stars that are not chemically stratified.
These elements include
germanium (Ge, $Z =$~32),
arsenic (As, $Z =$~33),
selenium (Se, $Z =$~34),
cadmium (Cd, $Z =$~48),
tellurium (Te, $Z =$~52),
lutetium (Lu, $Z =$~71),
tantalum (Ta, $Z =$~73),
tungsten (W, $Z =$~74),
rhenium (Re, $Z =$~75),
osmium (Os, $Z =$~76),
platinum (Pt, $Z =$~78), 
gold (Au, $Z =$~79), and
bismuth (Bi, $Z =$~83)
\citep{cowan96,cowan02,cowan05,
sneden98,sneden03,
roederer09,roederer10b,roederer12a,roederer12d,
barbuy11,peterson11,roederer12c,roederer12b,siqueiramello13,placco14}.
These studies have also
detected species not found in the optical domain,
such as
ionized molybdenum (Mo, $Z =$~42);
placed new upper limits on
tin (Sn, $Z =$~50) and
mercury (Hg, $Z =$~80); 
and increased the number of useful abundance indicators of
zirconium (Zr, $Z =$~40),
iridium (Ir, $Z =$~77), and
lead (Pb, $Z =$~82).
The high signal-to-noise (S/N) and high spectral-resolution requirements
for such studies limit \textit{HST} to observing nearby, NUV-bright stars,
so the number of stars where each of these heavy species have been studied
rarely exceeds $\sim$~5.

When found in low-metallicity stars,
these elements provide observational constraints
on the production of rare, heavy nuclei
in the earliest generations of stars.
They provide critical evidence
to locate the sites and constrain models of 
heavy-element nucleosynthesis via
charged-particle, \rpro, and \spro\ reactions.
For example,
in low-metallicity stars
the abundance of Ge
appears correlated with iron (Fe, $Z =$~26)
\citep{cowan05}.
Ge behaves like a primary element
whose production may be linked with Fe.
All previous observations of As and Se, 
however, only covered one star with
[Fe/H]~$< -$2 \citep{roederer12c,roederer12b}.
There is no way to evaluate from these data whether
As or Se correlate with Fe at low metallicity.
Strontium (Sr, $Z =$~38), 
the next heaviest element that is readily observable
in low-metallicity stars, is
not strongly correlated with Fe at low metallicity
(e.g., \citealt{luck85,mcwilliam98,travaglio04}).
In the earliest generations of stars,
these observations indicate that 
the lightest heavy nuclei whose
production is decoupled from 
iron-group nucleosynthesis
lie between Ge and Sr,
in the mass range 
72~$\leq A \leq$~88.

These studies also may
confirm line identifications
by corroborating abundances derived from multiple lines
of the same species.
The NUV domain of late-type stars
exhibits many unidentified absorption lines.
Abundances derived from one
absorption feature must be confirmed by others
to minimize the possibility
of misidentification.
\citet{barbuy11}, \citet{roederer12b}, and \citet{roederer12d} 
have made progress in this regard,
but some detections remain unconfirmed
(e.g., Cd~\textsc{i}, Te~\textsc{i}, 
Os~\textsc{ii}, and Ir~\textsc{ii}).

These studies also 
enable the detection of dominant ionization states 
(usually first ions)
of elements whose minority species are detected
in the optical domain.
Nickel (Ni, $Z =$~28), 
copper (Cu, $Z =$~29), 
zinc (Zn, $Z =$~30), 
and Mo are prominent examples
\citep{peterson11,roederer12b,wood14}.
These observations provide valuable constraints on possible systematic
effects, like over-ionization relative to the
local thermodynamic equilibrium (LTE) populations, 
that could impact the minority species.

\textit{HST} is currently the only source 
of high-resolution NUV spectra, 
so observing targets must be selected with care.
Our previous observations \citep{roederer12d}
revealed striking similarities between two 
physically-unrelated metal-poor giants, \hda\ and \hdb.
These stars are first-ascent red giants
with effective temperatures (\teff~$\approx$~5100~K) and 
surface gravities (\logg~$\approx$~2.6 in cgs units) 
that are identical within their uncertainties.
They have nearly identical
metallicities ([Fe/H]~$\approx -$2.5) 
and light-element abundances.
In contrast, the abundances of elements heavier than the 
iron-group in these two stars differ
by a factor of $\approx$~3.
Thus the spectra of these two stars
are virtually identical
except for the strength of lines of heavy elements.
This fortuitous combination of characteristics
simplifies the process
of searching for lines of heavy elements and
prompted us to acquire new 
STIS observations
extending farther into the NUV.~
Here we present these new observations.
We report new abundance data for several heavy elements
and new constraints on the sources of
systematic uncertainty discussed above.

Throughout this paper, we use
the standard definitions of elemental abundances and ratios.
For element X, the logarithmic abundance is defined
as the number of atoms of element X per 10$^{12}$ hydrogen atoms,
$\log\epsilon$(X)~$\equiv \log_{10}(N_{\rm X}/N_{\rm H}) +12.0$.
For elements X and Y, the logarithmic abundance ratio relative to the
solar ratio of X and Y is defined as
[X/Y]~$\equiv \log_{10} (N_{\rm X}/N_{\rm Y}) -
\log_{10} (N_{\rm X}/N_{\rm Y})_{\odot}$.
Abundances or ratios denoted with the ionization state
indicate the total elemental abundance as derived from 
that particular ionization state.
When reporting relative abundance ratios (e.g., [X/Fe]),
these ratios compare neutrals with neutrals and ions with ions
when possible.

\section{Observations}
\label{observations}

In program GO-12976, we obtained
new STIS observations \citep{kimble98,woodgate98}
of \hda\ and \hdb\ using
the E230M echelle grating, centered on 1978~\AA, and the
NUV Multianode Microchannel Array (MAMA) detector.
The 0\farcs06\,$\times$\,0\farcs2 slit 
yields a $\sim$~2~pixel 
resolving power (R~$\equiv \lambda/\Delta\lambda$) $\sim$~30,000.
This setup produces wavelength coverage from 1610--2365~\AA\ in a single
exposure,
although in practice only wavelengths longer than 
$\approx$~1900~\AA\ have
useful S/N ratios.

A total of 12 individual exposures of \hda\ were
scheduled over four visits
from 2013 May~21 to 2013 July~15,
resulting in a total exposure time of 28.6~ks.
Fifteen exposures of \hdb\ were scheduled over
three visits from 2013 Jan~22 to 2013 Feb~01,
resulting in a total exposure time of 37.4~ks.
The standard observational sequence includes
acquisition and peak-up images to center the star on the narrow slit.
These observations are reduced and 
calibrated using the \textit{calstis} pipeline.
After co-adding the individual exposures, the spectrum of each star has
S/N ratios ranging from 
$\sim$~10~pix$^{-1}$ near 1900~\AA\ to 
$\sim$~50~pix$^{-1}$ near 2100~\AA\ and 
$\gtrsim$~100~pix$^{-1}$ at 2350~\AA.

STIS observations of \hda\ and \hdb\ obtained previously
in program GO-12268 
covered the NUV spectral range from 2280--3115~\AA.
We co-add those observations with our new observations
in the region of overlap.
Final S/N ratios in the region from 2280--2365~\AA\ are
$\gtrsim$~130~pix$^{-1}$.

\section{Model atmospheres and stellar parameters}

We adopt the same ATLAS9 model atmospheres 
\citep{castelli04} used in our previous study.
For \hda, 
\teff, \logg,
microturbulent velocity (\vt), and metallicity ([M/H])
of the model atmosphere are 
5100~$\pm$~200~K, 2.67~$\pm$~0.2, 
1.50~$\pm$~0.2~\kmsec, and $-$2.37~$\pm$~0.20, 
respectively.
For \hdb, \teff, \logg, \vt, and [M/H] are 
5080~$\pm$~200~K, 2.57~$\pm$~0.2, 
1.60~$\pm$~0.2~\kmsec, and $-$2.46~$\pm$~0.22, 
respectively.

Here we briefly summarize the methods 
\citet{roederer12d} used to derive these parameters.
The \citet{alonso99} color-temperature relations
were used to calculate an initial estimate for
\teff\ from the dereddened $V-K$ color.
The \logg\ values were calculated from these temperatures
and \textit{Hipparcos} parallax measurements,
assuming a mass of 0.8~$\pm$~0.1~\msun.
Microturbulence velocities were derived
by minimizing correlations between
line strength and the abundances derived from 
optical Fe~\textsc{i} lines.
The initial \teff\ values were revised 
to minimize correlations between 
the excitation potentials (E.P.) 
and the abundances derived from optical Fe~\textsc{i} lines.
An analysis of interstellar absorption lines
toward each star supported the adjusted \teff\ values.
Revised \logg\ values were calculated accordingly.
Finally, the metallicity of the model atmosphere was 
adopted to be equivalent to the Fe abundance derived
from Fe~\textsc{ii} lines.
These steps were iterated until convergence
(\teff\ stable within 10~K, \logg\ within 0.01~dex,
\vt\ within 0.05~\kmsec, and [M/H] within 0.01~dex).

\section{Iron abundance trends with wavelength}

In our previous 
work on \hda\ and \hdb\ \citep{roederer12d}, 
we uncovered evidence that 
Fe~\textsc{i} lines did not yield consistent abundances
over the NUV and optical spectral ranges.
Specifically, abundances derived from 
lines with $\lambda <$~4000~\AA\
were lower than abundances derived from lines
with longer wavelengths.
The most severe deficiency in \hda\ and \hdb\ 
occurred for lines with 
3100~$< \lambda <$~3647~\AA, as illustrated in 
Figure~9 of \citeauthor{roederer12d} 
Subsequent studies by \citet{lawler13} and \citet{wood13}
of NUV and optical 
Fe~\textsc{i}, Ti~\textsc{i}, 
and Ti~\textsc{ii} lines
revealed similar inconsistencies
in the metal-poor subgiant 
\object[HD 84937]{HD~84937}.
However,
\citet{wood14} did not find any such 
inconsistencies in their analysis of Ni~\textsc{i} lines
in this star.
In an independent analysis of the metal-poor giant
\object[BD+44 493]{BD$+$44~493} by 
\citeauthor{ito13} (2013; see also \citealt{placco14}),
the Fe~\textsc{i} lines with $\lambda <$~3700~\AA\
yield a mean abundance about 0.15~dex lower than the
Fe~\textsc{i} lines with $\lambda >$~3700~\AA,
as shown by their Figure~6.

\begin{figure}
\begin{center}
\includegraphics[angle=0,width=3.2in]{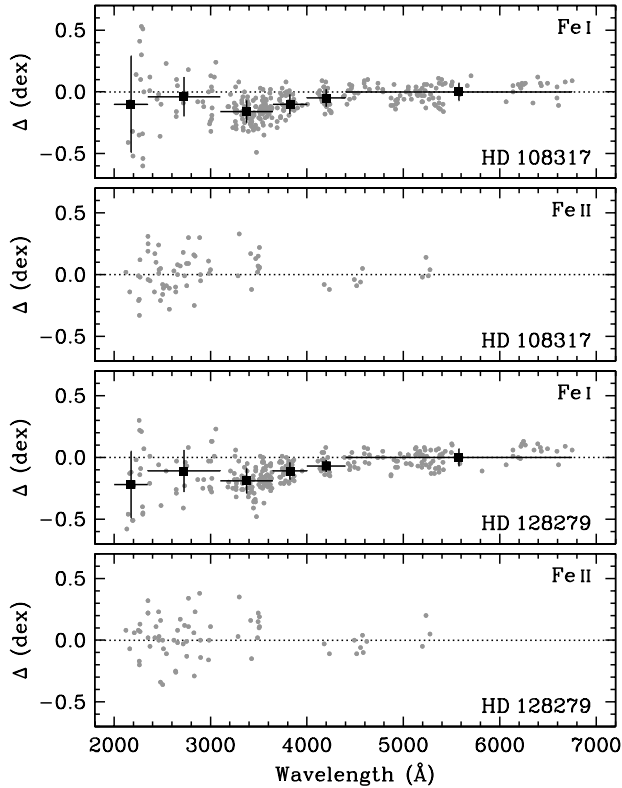}
\end{center}
\caption{
\label{waveplot1}
Relative Fe abundances from Fe~\textsc{i} and \textsc{ii} lines.
Gray dots represent abundances derived from individual lines.
Large black squares represent the mean abundance in
each wavelength range.
Horizontal error bars mark the wavelength range, and
vertical error bars indicate the standard deviation.
The dotted line in each panel marks the mean abundance 
derived from lines with $\lambda >$~4400~\AA.
}
\end{figure}

\begin{figure}
\begin{center}
\includegraphics[angle=0,width=3.2in]{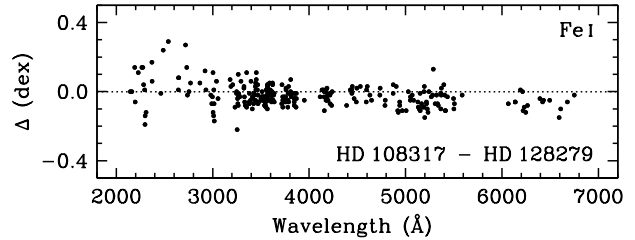}
\end{center}
\caption{
\label{waveplot2}
Line-by-line differences in the Fe~\textsc{i} 
abundances.
The differences are in the sense of 
\mbox{HD~108317} minus \mbox{HD~128279}.
The dotted line indicates a difference of zero.
These differences are independent
of the choice of \loggf\ values.
}
\end{figure}

These studies investigated several possible
causes of the abundance discrepancies.
The wavelength region most severely affected is that
around the transition from the Paschen continuum to the
Balmer continuum.
\citet{roederer12d} concluded that a missing source of
continuous opacity could account for part of the effect.
\citet{lawler13} and \citet{wood13,wood14}
suggested that departures from LTE could
also be partly responsible.
Missing opacity would impact all species with transitions
in the affected wavelength range equally, while non-LTE
effects would affect each transition individually, 
if at all.
Ultimately, \citeauthor{roederer12d}\ adopted an empirical set of 
wavelength-dependent abundance corrections for all transitions
based on the results from Fe~\textsc{i} lines.

Our new STIS data extend to even shorter wavelengths.
We measure equivalent widths of relatively unblended
Fe~\textsc{i} and Fe~\textsc{ii} lines in our data.
All \loggf\ values are adopted from the 
National Institute of Standards and Technology (NIST) 
Atomic Spectra Database (ASD; \citealt{kramida13}).
This critical compilation assesses the accuracy of these
\loggf\ values to be 0.03 to 0.30~dex
(grades B$+$ to D).
The wavelengths, E.P.,
\loggf\ values, estimated accuracies, 
and equivalent widths are presented in 
Table~\ref{ewtab}.

\input{tab1}

We derive the abundances
using a recent version of
the line analysis code MOOG \citep{sneden73} that includes the
contribution of Rayleigh scattering from H~\textsc{i} 
in the source function, as discussed by \citet{sobeck11}.
Figure~\ref{waveplot1} illustrates the abundances derived from lines of
Fe~\textsc{i} and Fe~\textsc{ii} as a function of wavelength.
New abundances for lines with 
$\lambda <$~2360~\AA\ are shown, together with the
data for lines at longer wavelengths
from \citet{roederer12d}.
All abundances in Figure~\ref{waveplot1} are 
referenced to the mean Fe abundance derived from
lines with $\lambda >$~4400~\AA.
The line-to-line dispersion
for Fe~\textsc{i} and Fe~\textsc{ii} lines at short wavelengths
in both stars
is significantly larger than for lines at longer wavelengths.
The abundance ``dip'' found by \citeauthor{roederer12d}\
for lines with 
3100~$< \lambda <$~4000~\AA\ 
exceeds the dispersion 
of Fe~\textsc{i} at these wavelengths.
No dip is apparent for Fe~\textsc{ii} lines, 
but the number of lines in the affected region is small.

To eliminate the possibility 
that uncertainties in the \loggf\ values
are a source of this error,
we show the 
line-by-line differential abundances
between \hda\ and \hdb\ in Figure~\ref{waveplot2}.
For Fe~\textsc{i}, the dispersion for lines with $\lambda <$~3100~\AA,
0.093~dex, is two times larger than the 
dispersion for lines with $\lambda >$~3500~\AA,
0.045~dex.
For Fe~\textsc{ii}, the dispersion for lines with $\lambda <$~3100~\AA,
0.090~dex, is also several times larger than the
dispersion for lines with $\lambda >$~3500~\AA,
0.034~dex.
The S/N ratios are $\gg$~100 for the STIS, HIRES, and MIKE
spectra used from 2350~\AA\ to 6800~\AA,
and the S/N ratios are $>$~50 from 2100~\AA\ to 2350~\AA.
This demonstrates the challenge of deriving 
reliable abundances from crowded regions of spectra.
This also suggests the presence of a threshold in abundance precision
from being limited by S/N to being limited by
line blends, continuum placement, etc.
In principle, significantly higher spectral resolution
($R \gtrsim$~60,000)
would help to identify the continuum level and
signal the presence of blending lines.

Lacking further information about how to properly treat
each species of each element, 
we follow \citet{roederer12d} in adopting a set of 
wavelength-dependent abundance corrections for all transitions
based on the results from Fe~\textsc{i} lines.
The effect of these corrections is to adopt a 
local metallicity in hopes of preserving 
accurate ratios of one element to another.
Our adopted corrections are listed in Table~\ref{corrtab}.
For completeness, Table~\ref{corrtab} 
also lists the corrections for longer wavelengths 
from Table~10 of \citeauthor{roederer12d} 

\input{tab2}

\section{Heavy element line detection}

\begin{figure*}
\begin{center}
\includegraphics[angle=0,width=3.0in]{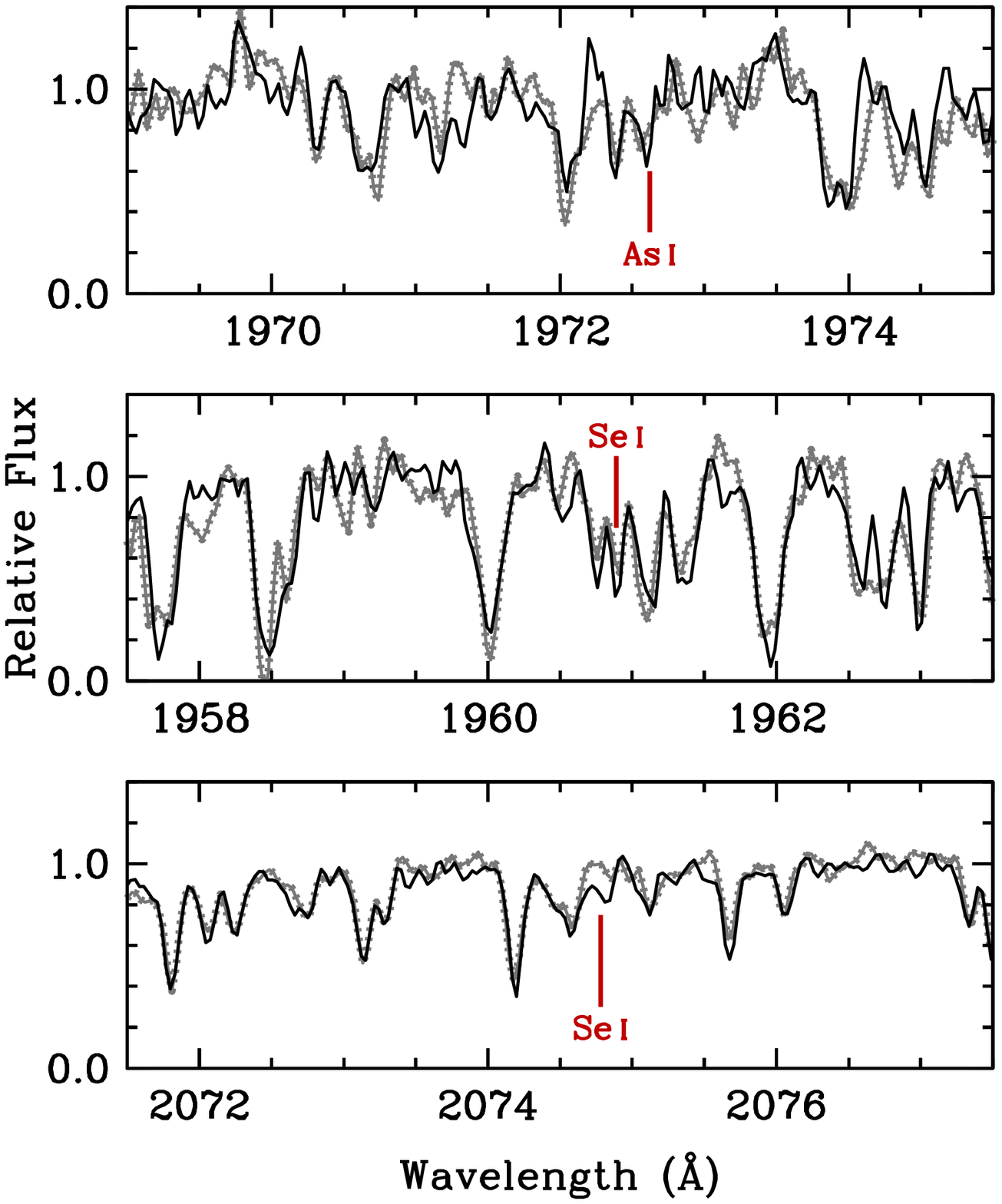}
\hspace*{0.3in}
\includegraphics[angle=0,width=3.0in]{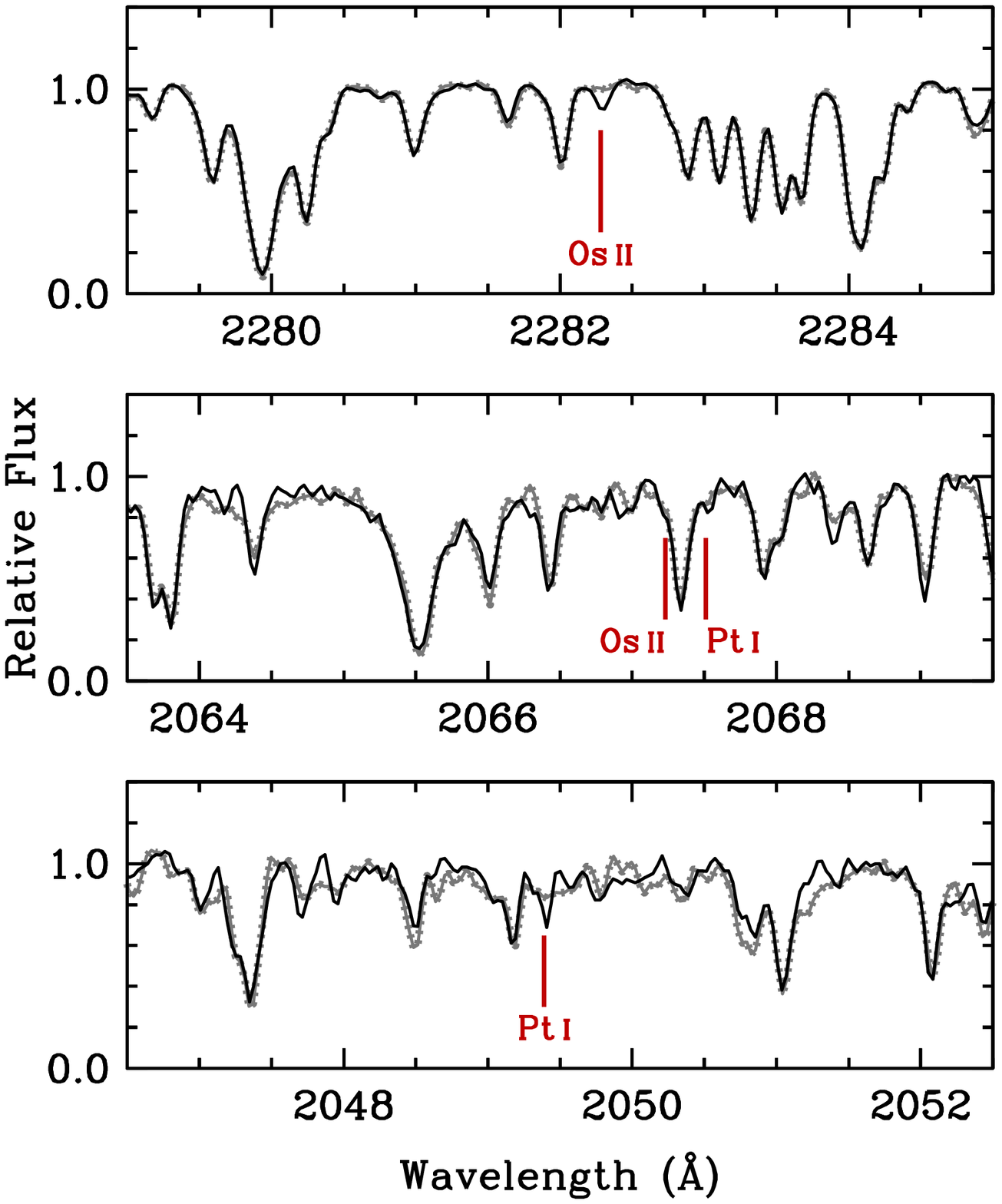}
\end{center}
\caption{
\label{specplot2}
The spectra of \mbox{HD~108317} (thin black curve)
and \mbox{HD~128279} (studded gray curve)
around several lines of interest.
Air wavelengths are shown for data with $\lambda >$~2000~\AA,
and vacuum wavelengths are shown for data with $\lambda <$~2000~\AA.
Note the similarity between the spectra of these two
stars except at the wavelengths of absorption features
arising from heavy elements.
}
\end{figure*}

\begin{figure*}
\begin{center}
\includegraphics[angle=0,width=3.0in]{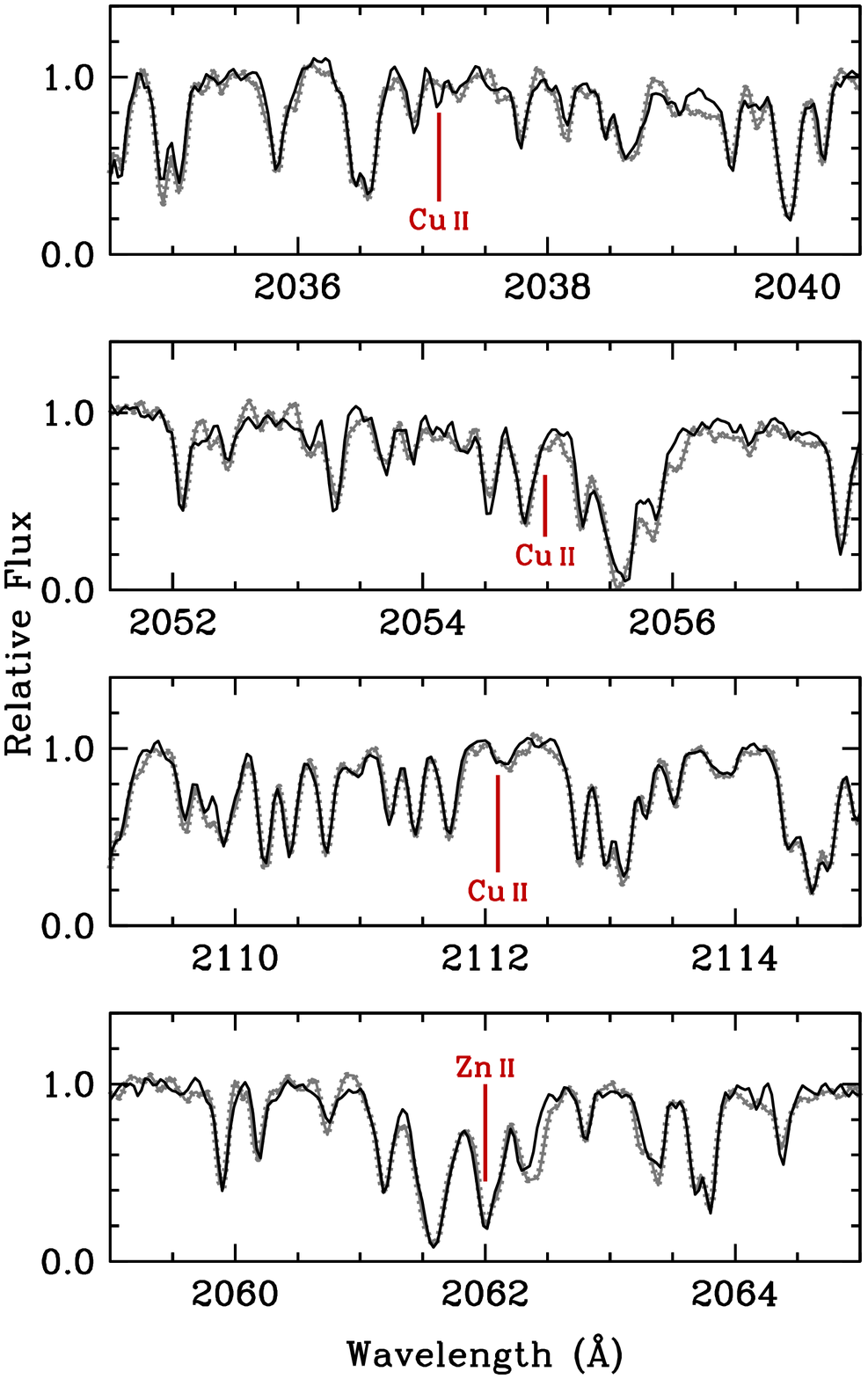}
\hspace*{0.3in}
\includegraphics[angle=0,width=3.0in]{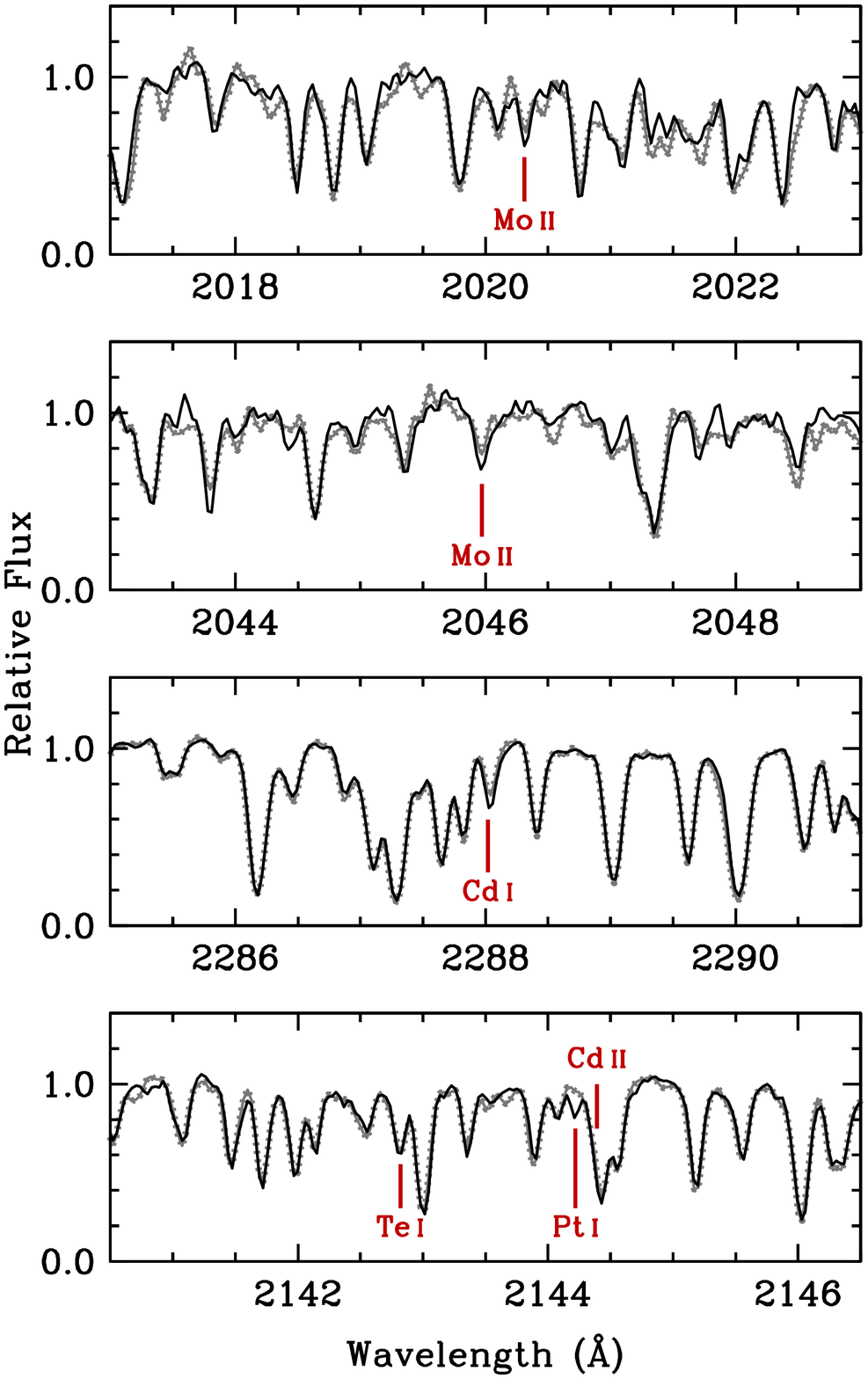}
\end{center}
\caption{
\label{specplot1}
The spectra of \mbox{HD~108317} (thin black curve)
and \mbox{HD~128279} (studded gray curve)
around several lines of interest.
As in Figure~\ref{specplot2},
note the similarity between the spectra of these two
stars except at the wavelengths of absorption features
arising from heavy elements.
}
\end{figure*}

We follow the same approach as
\citet{roederer12d} to identify
reliable abundance indicators of heavy elements
in the NUV spectra of \hda\ and \hdb.
We search for transitions reported in recent studies by
\citet{peterson11},
\citet{roederer12b}, 
\citet{roederer12d}, and
\citet{siqueiramello13}.
We search for leading lines in the NUV
that are listed in recent laboratory or theoretical studies
of heavy element spectra
\citep{denhartog05,nilsson08}.
We also overplot the high S/N spectra of \hda\ and \hdb\
and search for significant differences by eye.
As demonstrated by 
\citet{roederer12a,roederer12d},
these differences frequently indicate the presence 
of heavy element absorption lines in \hda.
We then generate a synthetic spectrum for each line.
Many candidates are too weak or too blended
to yield a secure detection.
Figures~\ref{specplot2} and \ref{specplot1}
illustrate sections of the STIS spectra 
around 17~lines that are detected in 
one or both stars.

\section{Abundance analysis of heavy elements}

The atomic data for lines of interest
are listed in Table~\ref{atomictab}.
References and uncertainties
for the \loggf\ values are noted.
We use MOOG to generate synthetic spectra, and
we derive abundances or upper limits by spectrum-synthesis matching.
We include hyperfine splitting (hfs) 
of As~\textsc{i}, Cd~\textsc{ii}, Yb~\textsc{ii}, and Hg~\textsc{ii}.
Isotope shifts (IS)
are included in our syntheses
of Cd~\textsc{ii}, Yb~\textsc{ii}, and Hg~\textsc{ii},
and we use the \rpro\ isotope mixtures presented by \citet{sneden08}.
The complete hfs and IS line component patterns
for these transitions are presented in Tables~4, 8, 10, and 12
of \citet{roederer12b}.

\input{tab3}

\begin{figure*}
\begin{center}
\includegraphics[angle=0,width=3.0in]{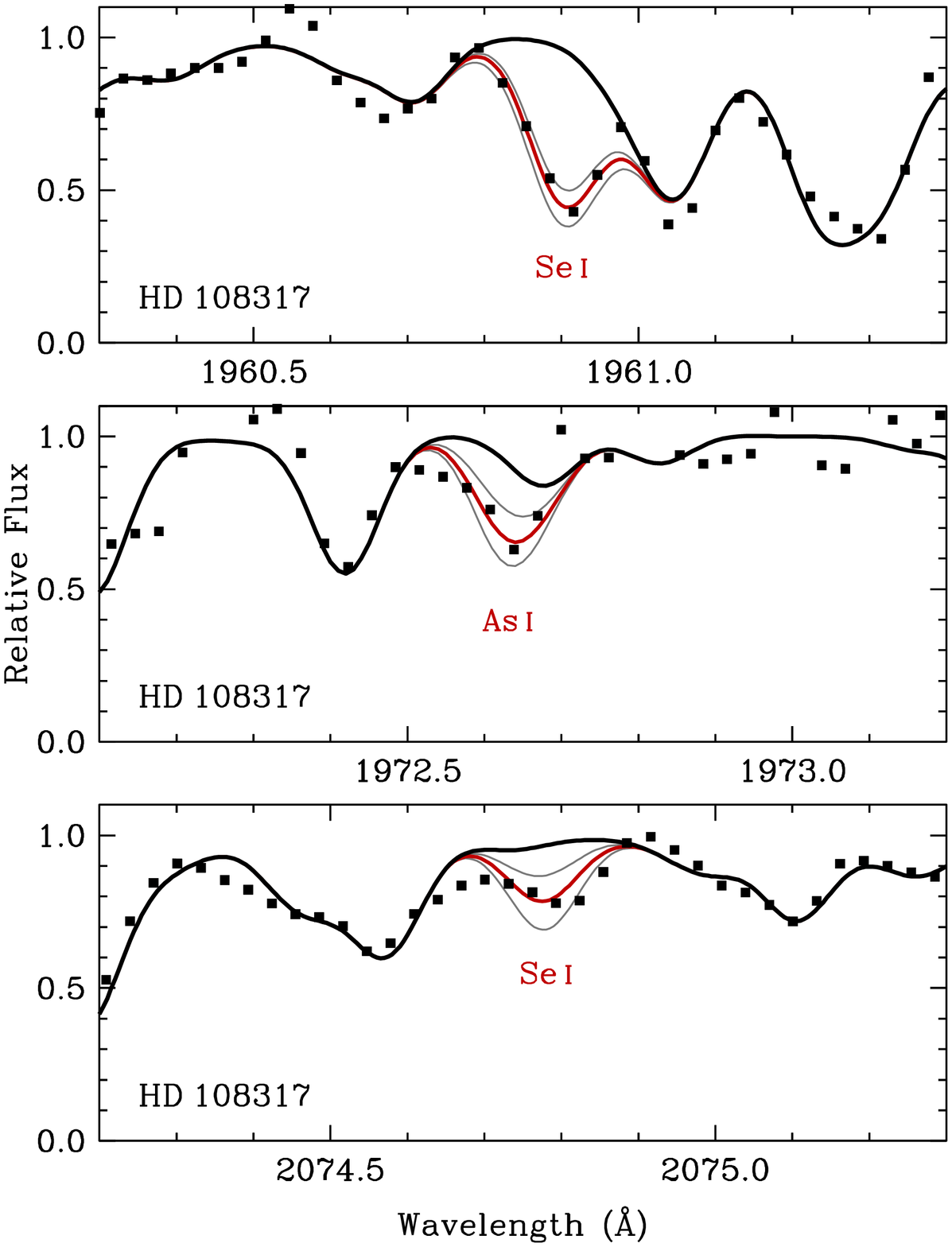}
\hspace*{0.3in}
\includegraphics[angle=0,width=3.0in]{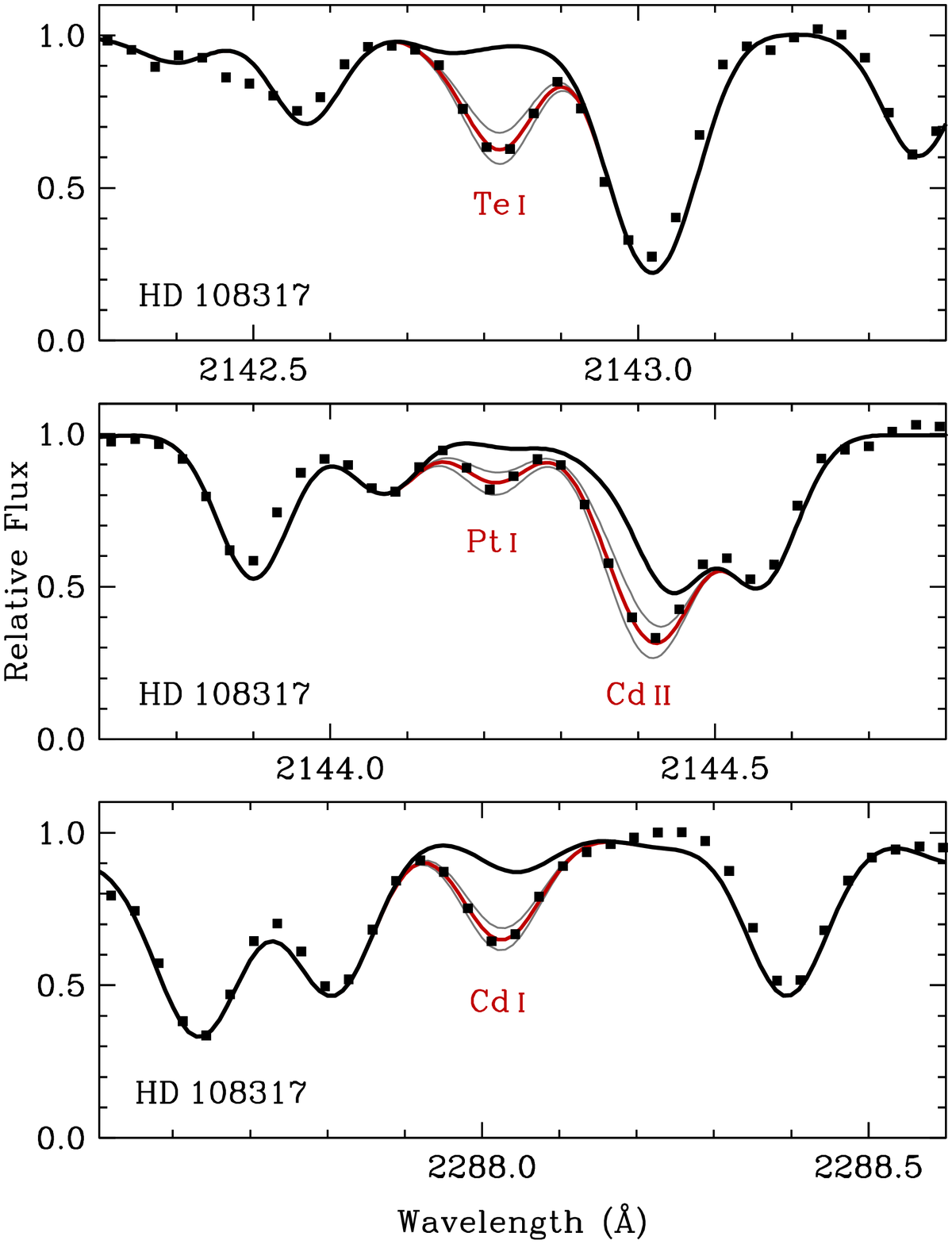}
\end{center}
\caption{
\label{synplot}
Comparison of synthetic and observed spectra around 
several lines of interest in \mbox{HD~108317}.
The observed spectrum is indicated by the filled squares.
The best-fit synthesis is indicated by the thick red line.
Variations in the best fit are shown by the thin gray lines;
these are 
$\pm$~0.2~dex (Cd~\textsc{i}, Pt~\textsc{i}),
$\pm$~0.3~dex (Cd~\textsc{ii}, Te~\textsc{i}), and 
$\pm$~0.4~dex (As~\textsc{i}, Se~\textsc{i}).
The thick black lines represent a syntheses with no
absorption from the line(s) of interest.
}
\end{figure*}

Several examples of key lines of interest in \hda\
are illustrated in Figure~\ref{synplot}.
Our procedure for generating the line lists and producing
acceptable matches to the observed spectra is
described in detail in \citet{roederer12d}.
We briefly summarize these methods here.
There are numerous 
unidentified lines in NUV spectra of
low-metallicity late-type stars, and
the transition probabilities for many 
identified lines are not known
with sufficient accuracy
to produce acceptable 
matches between the observed and synthetic spectra
without additional effort.  
Transition probabilities measured by laboratory experiments
are used in our syntheses whenever possible.
Unidentified lines are fit assuming the
absorption arises from Fe~\textsc{i} lines
with a lower E.P.\ of 1.5~eV,
and we use the observed line profile to
constrain the strength of each line.
Lines of interest must be
distinct from such lines 
to merit use as abundance indicators.

\section{Comments on individual elements}

In this section, we discuss the 
detection of multiple lines
as a way to confirm the line identifications
for Te, Ir, and Pt.
We also discuss the limitations 
of line identification for Pb and Bi.

\subsection{Tellurium}

Although two Te~\textsc{i} lines have been used as 
abundance indicators previously,
the spectra of the four stars studied by
\citet{roederer12b} and \citet{roederer12a} 
did not cover both lines simultaneously
in any of these stars.
Now, our combined NUV observations of \hda\ and \hdb\
cover both Te~\textsc{i} lines (2142 and 2385~\AA)
in these two stars.
The abundances derived from these lines agree for each star.
In \hda\ the Te~\textsc{i} 2142 and 2385~\AA\ lines yield 
$\log \epsilon$~(Te)~$= +$0.14~$\pm$~0.30 and
$+$0.02~$\pm$~0.25, and 
in \hdb\ these lines yield 
$\log \epsilon$~(Te)~$= -$0.04~$\pm$~0.30 and
$-$0.18~$\pm$~0.25.
This increases our confidence that both detections are legitimate 
and that both lines are reliable abundance indicators.

\subsection{Iridium}

A line of Ir~\textsc{ii} at 2126~\AA\ has been used previously by
\citet{roederer12b} as an abundance indicator in \hdc.
This line also appears in our spectra of \hda\ and \hdb.
Its presence in \hdb\ is surprising, since no absorption lines
of the neighboring elements Os or Pt
are detected in \hdb.
If we assume that the absorption at 2126.81~\AA\
in \hdb\ is due to Ir~\textsc{ii}, we derive an
Ir abundance ($\log \epsilon \approx -$0.3~$\pm$~0.3) 
that is 1.0~dex higher than the upper limit on Os
($\log \epsilon < -$1.28) and
0.5~dex higher than the upper limit on Pt
($\log \epsilon < -$0.78).
This suggests that the absorption at 2126.81~\AA\ 
is probably not due to Ir~\textsc{ii}
in \hda\ or \hdb.
Instead, this absorption can be fit by
increasing the \loggf\ value of the Fe~\textsc{i}
line at 2126.82~\AA.
Conservatively, we assume that the absorption 
detected in \hdc\ is also
not due to Ir~\textsc{ii}.
The Ir abundances reported in Tables~14 and 15
of \citeauthor{roederer12b} should be regarded as upper limits.

\subsection{Platinum}

\citet{roederer12b} derived Pt abundances from
the Pt~\textsc{i} 2049.39 and 2067.51~\AA\ lines 
in \hdc,
and we also detect these lines in \hda.
One new line of Pt~\textsc{i}, 2144.21~\AA, is also 
detected in \hda\ and
listed in Table~\ref{atomictab}.
\citet{denhartog05} measured the radiative
lifetime of the upper level 
of this transition
[$5d^{8}6s6p$ ($^{2}F$)$^{3}D$; 
46622.489~cm$^{-1}$].
The 2144.21~\AA\ line is a dominant
(68\%) decay branch to the ground level
($4d^{9}6s$ $^{3}D_{3}$).
We use these measurements to calculate the \loggf\ value
of this line, $-$0.37~$\pm$~0.07.
The abundances derived from all three Pt~\textsc{i}
lines are in agreement, 
as shown in Table~\ref{atomictab}.
They are also in agreement with the abundances derived from
two Pt~\textsc{i} lines at longer wavelengths in \hda,
listed in Table~3 of \citet{roederer12d}.

\subsection{Lead}

Only a few lines of neutral Pb have been detected in
the optical and NUV spectral range,
and these are usually blended with other features
that complicate attempts to secure a reliable detection.
Neutral Pb never comprises more than a few
percent of all Pb atoms 
in the conditions found in late-type stellar atmospheres
\citep{mashonkina12,roederer12b}.
\citeauthor{mashonkina12}\ investigated the effect of
overionization on Pb~\textsc{i} lines and presented a
set of non-LTE corrections for abundances derived from Pb~\textsc{i} lines.
For the Pb~\textsc{i} 4057~\AA\ line in \hda,
their calculations suggest the correction is $+$0.52~dex.

Ionized Pb is even more elusive.
The NIST database lists only
eight classified Pb~\textsc{ii} transitions
from 2000--10000~\AA, and
only one of these transitions 
originates from an electronic level 
below 7~eV.
The 2203.53~\AA\ transition connects the upper 
$6s^{2}7s$ $^{2}S_{1/2}$ level to the lower
$6s^{2}6p$ $^{2}P^{o}_{3/2}$ level.
It is difficult to 
estimate the accuracy of the \loggf\ value
of this line,
which \citet{morton00} reports
based on calculations by \citet{migdalek83}.

\begin{figure}
\begin{center}
\includegraphics[angle=0,width=3.2in]{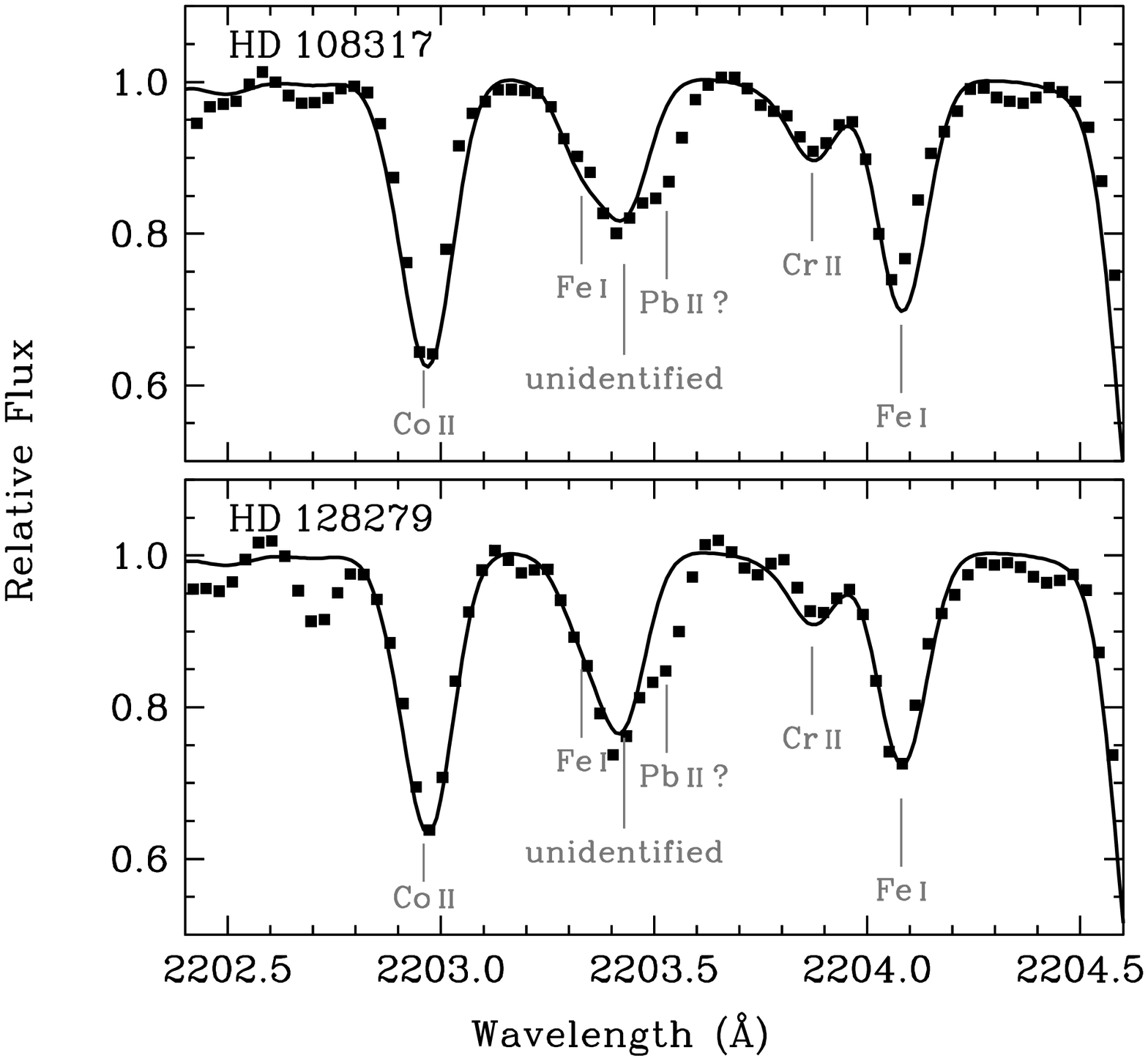}
\end{center}
\caption{
\label{pbplot}
Comparison of observed and synthetic spectra around the
Pb~\textsc{ii} line in \mbox{HD~108317} and \mbox{HD~128279}.
The filled squares mark the observed spectra, and the curves
indicate a synthesis with no Pb~\textsc{ii} absorption.
To facilitate comparisons,
the label locations on the plots are
identical in the top and bottom panels.
}
\end{figure}

A small amount of absorption 
is detected at the predicted wavelength of the 
Pb~\textsc{ii} line, 2203.53~\AA.
As shown in Figure~\ref{pbplot}, this absorption falls
on the red wing of a stronger, unidentified feature with
wavelength 2203.44~\AA.
A weak Fe~\textsc{i} line also shows absorption in the blue
wing of this unidentified feature.
We generate a fictitious line at 2203.44~\AA\ in our synthesis
to reproduce the observed line profile in both
\hda\ and \hdb, yet residual absorption
at the predicted wavelength of the Pb~\textsc{ii} line
persists.
As shown in Figure~\ref{pbplot}, this absorption 
is present in both \hda\ and \hdb.
\citet{roederer12d} found that
none of the
elements at or beyond the \third\ \rpro\ 
peak could be detected in \hdb.
It would be uncharacteristic of the heavy-element abundance 
patterns in these two stars if absorption due to Pb~\textsc{ii}
was of nearly equal strength in each one.
There are many unidentified absorption features in 
the NUV spectra of these two stars, and 
we suspect that the residual absorption 
at 2203.53~\AA\ is probably not due to Pb~\textsc{ii}
in \hda\ or \hdb.
We derive an upper limit on the Pb abundance
from this Pb~\textsc{ii} line assuming all of the residual
absorption is due to Pb~\textsc{ii}.

\subsection{Bismuth}

\begin{figure}
\begin{center}
\includegraphics[angle=0,width=3.2in]{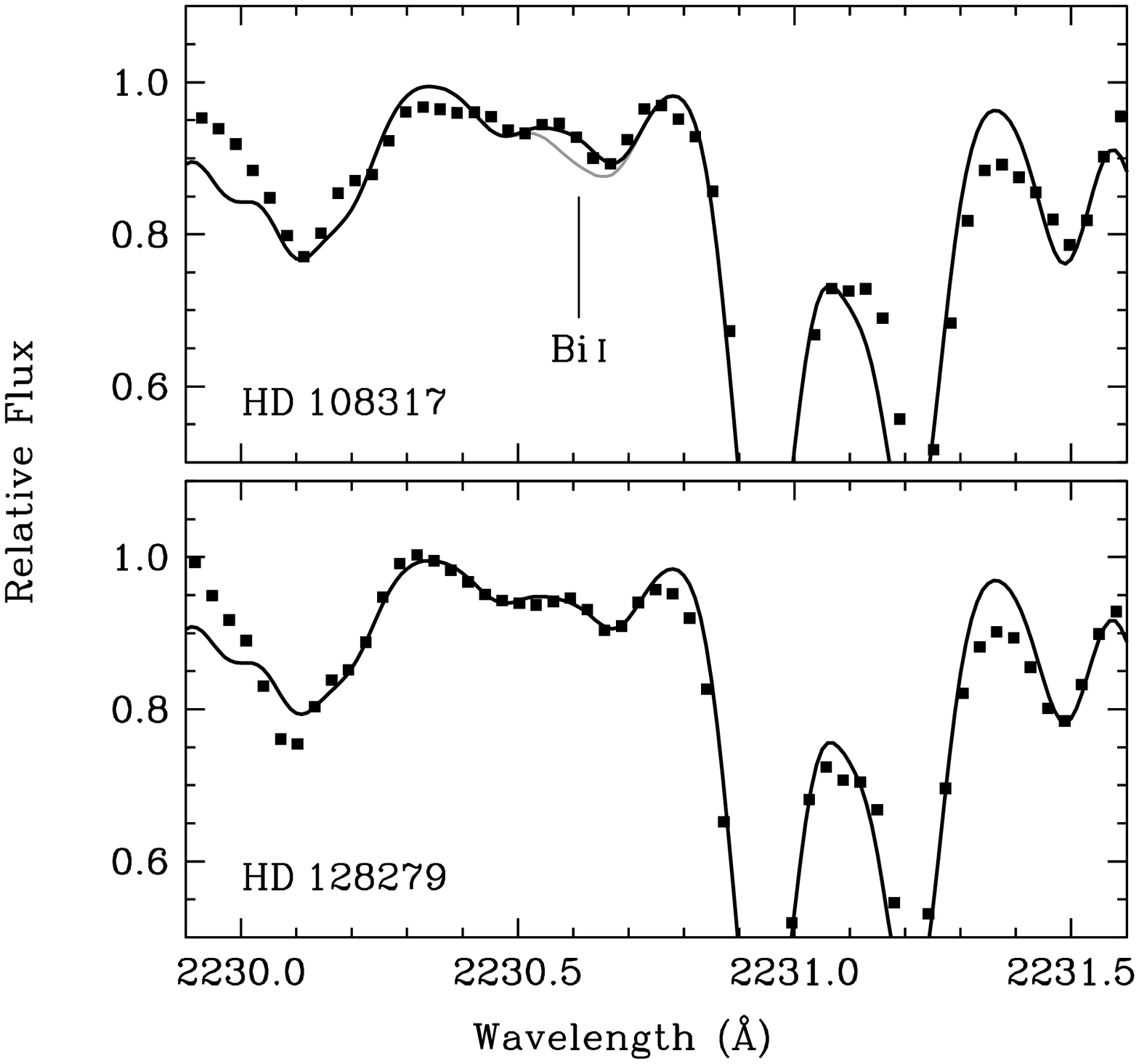}
\end{center}
\caption{
\label{biplot}
Comparison of observed and synthetic spectra around the
Bi~\textsc{i} line in \mbox{HD~108317} and \mbox{HD~128279}.
The filled squares mark the observed spectra.
The black lines
indicate a synthesis with no Bi~\textsc{i} absorption, and
the gray line in the top panel indicates our upper limit
on the Bi abundance.
}
\end{figure}

We derive an upper limit on the Bi
abundance from the Bi~\textsc{i} resonance line at 2230.61~\AA\ in \hda.
As shown in Figure~\ref{biplot}, this feature is weak, but
\citet{morton00} indicates that it should
be among the strongest of all Bi~\textsc{i} lines in late-type stars.
Other species may also contribute to the absorption at this 
wavelength. 
To estimate the contribution from Bi~\textsc{i}, 
we employ a technique we have used previously to 
estimate the absorption due to Au in \hda\ 
(see Appendix~A of \citealt{roederer12d}).
Since the \third\ \rpro\ peak elements are not detected in \hdb,
we assume that all absorption at this wavelength in \hdb\ is not 
due to Bi, and we adjust our line list to fit the observed line profile
in \hdb.
We then use this line list to fit the absorption features in \hda.
The fit is not satisfactory, so we cannot 
state with confidence that the Bi~\textsc{i} line is detected.
Figure~\ref{biplot} illustrates our derived upper limit.

\section{Results}

Table~\ref{atomictab} lists the
abundances or upper limits derived from each line.
Table~\ref{abundtab} lists the mean abundances or upper limits
for each species.
The means include lines from 
other NUV transitions at longer wavelengths,
when available.
Values in Table~\ref{abundtab} supersede those
published previously by \citet{roederer12a,roederer12d}.

\subsection{Summary of statistical uncertainties}

The means in Table~\ref{abundtab} are 
weighted by the statistical uncertainty 
obtained from the quadrature sum of
the fitting uncertainty (Table~\ref{atomictab}),
uncertainty in the \loggf\ values (Table~\ref{atomictab}),
and the uncertainty introduced by the wavelength-dependent corrections
(Table~\ref{corrtab}).
The uncertainties in Table~\ref{atomictab} reflect 
fitting uncertainties
estimated during the spectrum-synthesis matching.
The main sources of error are continuum placement
and blending features whose strength cannot be
empirically set by fitting the observed line profile.
These uncertainties are typically larger than 
those estimated for optical data.

\input{tab4}

\subsection{Systematic uncertainties}

Systematic uncertainties resulting from
errors in the model atmosphere parameters
have been discussed previously \citep{cowan05}, 
and we adopt their assessment of the uncertainties.
This uncertainty amounts to 0.12~dex for lines of neutral species
and 0.17~dex for lines of ionized species.
These uncertainties are added in quadrature with the statistical errors
listed in Table~\ref{abundtab} to 
form the total errors listed there.

We detect multiple transitions of
many of the species examined in our study.
For Cu~\textsc{ii}, 
Ge~\textsc{i}, Se~\textsc{i}, Mo~\textsc{ii},
Te~\textsc{i}, and Pt~\textsc{i}, 
the abundances derived from multiple transitions
are in good agreement.
These values are listed in Table~\ref{atomictab}
in the present study and Table~3 of \citet{roederer12d}.

We detect neutral and singly-ionized states of
Cu, Zn, Mo, Cd, and Os.
The neutral species of Cu, Zn, and Mo
were detected from optical transitions by \citet{roederer12d}.
These abundances are listed in Table~\ref{iontab}.
The [Zn/Fe], [Mo/Fe], [Cd/Fe], and [Os/Fe] ratios derived from
neutral and ionized atoms agree, although
the uncertainties are substantial in some cases.
Nevertheless, this places limits on the
magnitude of overionization or other non-LTE effects 
that could occur for the transitions studied in these two stars.

\input{tab5}

In contrast, the [Cu/Fe] ratio derived from ions is 
higher than that derived from the neutral species
in \hda\ and \hdb:\
$\langle$[Cu~\textsc{ii}/Cu~\textsc{i}]$\rangle = +$0.56~$\pm$~0.23.
For completeness,
we note that \citet{roederer12b} found a difference of 
$+$0.29~$\pm$~0.35~dex
in the [Cu/Fe] ratios derived from Cu~\textsc{ii} and \textsc{i}
lines in the warm (5950~K) metal-poor ([Fe/H]~$= -$1.8) 
subgiant \hdc.
Most Cu atoms are ionized in these stellar atmospheres,
so overionization of Cu~\textsc{i} relative
to the LTE populations could occur.
The Cu~\textsc{i} abundances in these three stars
are derived from the 3247 and 3273~\AA\ resonance lines.
\citet{bonifacio10} found that these lines did not give
consistent results with the Cu~\textsc{i}
5105 and 5782~\AA\ lines
in dwarfs and giants in two metal-poor globular clusters.
Their 3D LTE modeling of the line formation could
not fully reconcile the abundances,
and it was not apparent which set of lines (if either)
could be reliable Cu abundance indicators
given the current state of modeling tools.
Our results further suggest that
Cu~\textsc{i} lines may not be formed in LTE,
but clearly more work is needed to better understand
the formation of Cu lines in cool stars.

\citet{roederer12d} 
discussed the possible
discrepancy between the [Os/Fe] ratios
derived from Os~\textsc{i} and \textsc{ii} lines in \hda:\
the Os~\textsc{i} 3058~\AA\ line
yields an abundance
([Os/Fe]~$=+$0.63~$\pm$~0.23)
higher than that given by 
the Os~\textsc{ii} 2282~\AA\ line 
([Os/Fe]~$=+$0.14~$\pm$~0.42).
\citet{roederer10b} reported an abundance difference
of $+$0.34~$\pm$~0.30~dex between the Os~\textsc{i} 3058~\AA\ line and
the Os~\textsc{ii} 2282~\AA\ line in the metal-poor giant \bd.
Each difference is not significant given the uncertainties,
and other results suggest these differences are
not symptoms of systematic uncertainties.
\citet{cowan05} found that the Os~\textsc{i} 3058~\AA\ line
gave concordant results with the Os~\textsc{i} 2838 and 3301~\AA\ lines
($\sigma =$~0.08~dex) in \bd.
\citet{ivarsson04}\ found that the Os~\textsc{ii} 2282~\AA\ line
also gave concordant results with the Os~\textsc{ii} 2067
and 2070~\AA\ lines ($\sigma =$~0.1~dex) 
in the chemically peculiar star 
\object[chi Lup]{$\chi$~Lupi}.
The extant data do not indicate systematic differences
between the Os abundance indicators,
but reducing the statistical uncertainties and
examining abundances derived from these lines in additional stars
would provide worthwhile checks.

As shown in Table~\ref{iontab},
the upper limits on [Pb/Fe] derived from
the Pb~\textsc{ii} 2203~\AA\ line are slightly
higher than the strongest upper limits on [Pb/Fe]
derived from the Pb~\textsc{i} 2833~\AA\ line
\citep{roederer12d}.
\citet{roederer10c} reported a detection of the 
Pb~\textsc{i} 4057~\AA\ line in \hda\ based on a 
higher-quality spectrum taken with the
Tull Spectrograph \citep{tull95} on the
Smith Telescope at McDonald Observatory.
The line is weak, and that detection is tenuous.
If it is indeed a detection, however,
we would derive [Pb/Fe]~$\approx +$0.1~$\pm$~0.3
using our model atmosphere for \hda.
The \citet{mashonkina12} non-LTE correction to the Pb abundance
derived from the Pb~\textsc{i} 4057~\AA\ line implies
[Pb/Fe]~$\approx +$0.62~$\pm$~0.3.
If the [Pb/Fe] derived in LTE from Pb~\textsc{ii} lines
($<+$0.42)
accurately reflects the Pb abundance, then 
our upper limit is $\approx$~0.2~dex lower than the
value predicted by the non-LTE corrections.
This difference is within the uncertainties,
so it does not justify revision of the non-LTE corrections
or rejection of the detection.

\section{Discussion}
\label{discussion}

\subsection{The [As/Fe] and [Se/Fe] ratios}
\label{gce}

\begin{figure}
\begin{center}
\includegraphics[angle=0,width=3.2in]{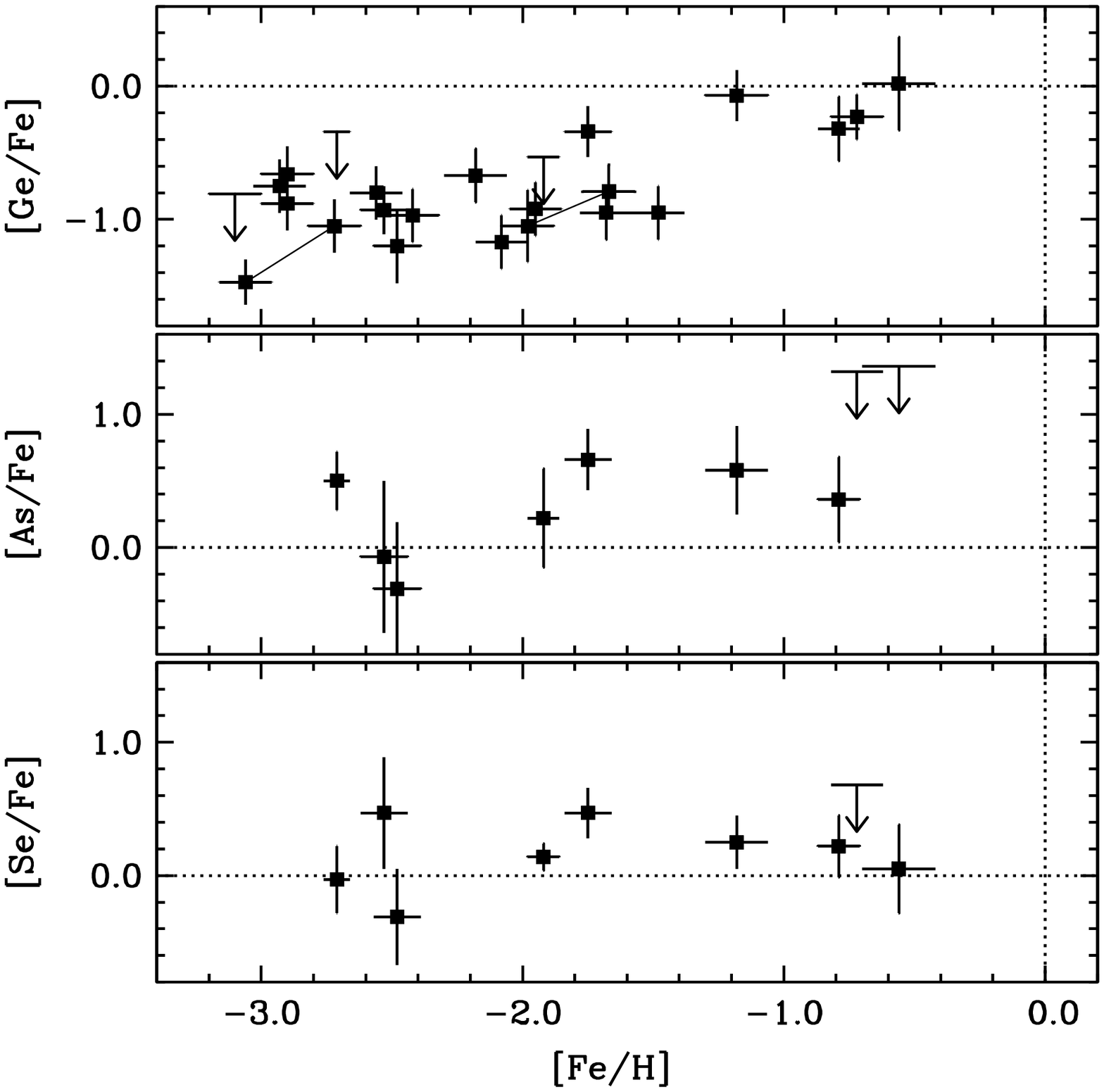}
\end{center}
\caption{
\label{xfeplot}
[Ge/Fe], [As/Fe], and [Se/Fe] ratios
for all late-type
stars with reported Ge, As, or Se abundances.
The dotted lines represent the solar ratios.
The [Fe/H] values shown are derived from optical Fe~\textsc{i} lines.
This figure includes abundances from 
\citet{sneden98},
\citet{cowan02,cowan05},
\citet{hill02},
\citet{ivans06},
\citet{roederer12c},
\citet{roederer12b},
\citet{roederer12d},
\citet{siqueiramello13},
and the present study.
The thin lines in the [Ge/Fe] panel connect the abundances reported
by \citet{roederer12d} to \citet{cowan96,cowan05} or \citet{sneden98} 
for \mbox{HD~122563} and \mbox{HD~126238}.
All [Ge/Fe] ratios have been corrected to the NIST \loggf\
abundance scale.
}
\end{figure}

Figure~\ref{xfeplot} illustrates the
[As/Fe] and [Se/Fe] ratios
in \hda, \hdb, and seven other stars
where As and Se have been studied.
The data show no compelling evidence for a cosmic dispersion
in either [As/Fe] or [Se/Fe].
On average, both ratios are
slightly super-solar 
over the entire metallicity range considered.
The mean [As/Fe] and [Se/Fe] ratios 
are $+$0.28~$\pm$~0.14 ($\sigma =$~0.36~dex)
and $+$0.16~$\pm$~0.09 ($\sigma =$~0.26~dex),
respectively.
In these expressions of uncertainties, 
the former value (e.g., 0.14~dex) is the standard error of the mean,
and the latter value (e.g., 0.36~dex) is the sample standard deviation.

Figure~\ref{xfeplot} also illustrates the 
[Ge/Fe] ratios in \hda, \hdb, and
17~other stars where Ge has been detected.
There is relatively little change in the
[Ge/Fe] ratio for stars with metallicities in the range 
$-$3.1~$<$~[Fe/H]~$< -$1.4, a fact
first noticed by \citet{cowan05}.
That study also noticed that 
[Ge/Fe] showed no correlation with 
[Eu/Fe] for the 10~stars in their sample.
\citeauthor{cowan05}\ concluded 
that Ge production in the early Galaxy
was decoupled from the \rpro\ nucleosynthesis
responsible for heavy elements, such as europium
(Eu, $Z =$~63).

In contrast with the [As/Fe] and [Se/Fe] ratios,
the mean [Ge/Fe] ratio in 
stars with [Fe/H]~$< -$1.4 is
$-$0.91~$\pm$~0.07 ($\sigma =$~0.26~dex).
At higher metallicities, the [Ge/Fe] ratio
increases toward the solar ratio,
whereas [As/Fe] and [Se/Fe] do not show a similar increase.
\citet{roederer12c} tentatively attributed 
the increase in [Ge/Fe]
to the weak component of the \spro\
(see also \citealt{pignatari10}),
which does not appear to have contributed
noticeable amounts of As or Se to these stars.

\subsection{The heavy element abundance patterns}
\label{abundpatterns}

\begin{figure*}
\begin{center}
\includegraphics[angle=270,width=3.0in]{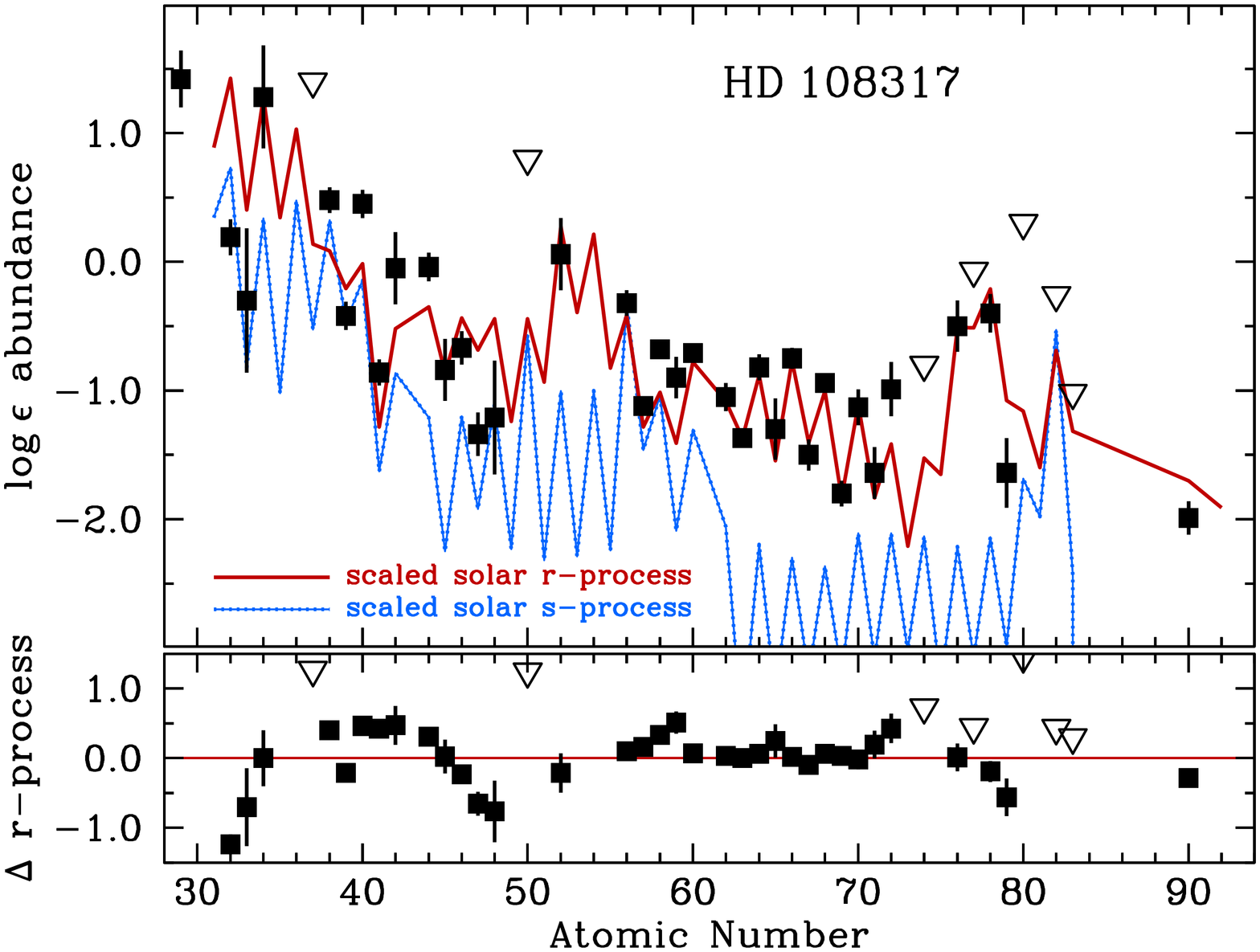}
\hspace*{0.3in}
\includegraphics[angle=270,width=3.0in]{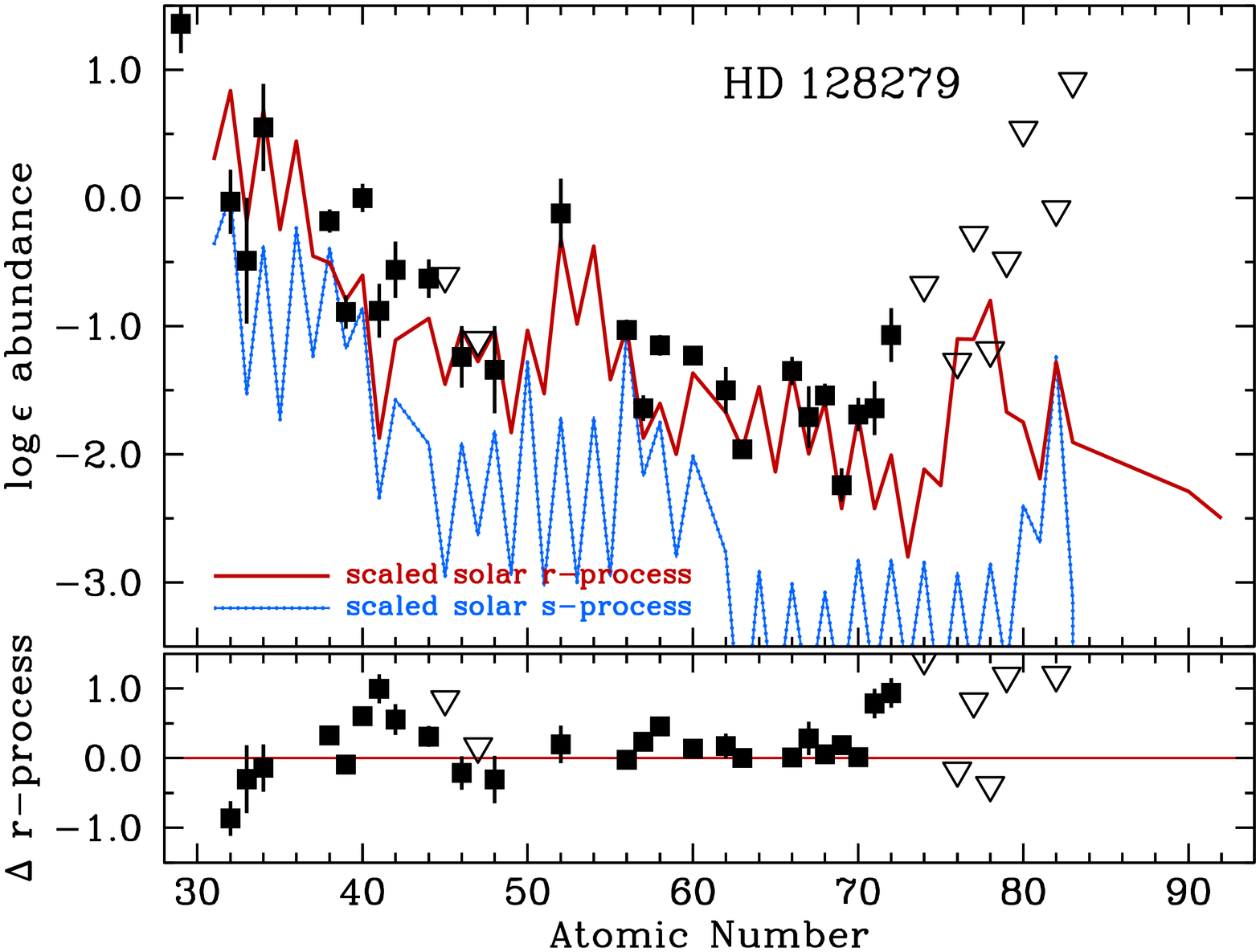}
\end{center}
\caption{
\label{logeplot1}
The heavy-element abundance patterns in
\mbox{HD~108317} and \mbox{HD~128279}.
These abundances are reported in 
Table~\ref{abundtab} of the present study
and Table~7 of \citet{roederer12d}.
Filled squares indicate detections, and 
open downward-pointing triangles indicate upper limits.
The solid red lines represent the 
Solar System \rpro\ pattern, normalized to Eu,
and the studded blue lines represent the 
Solar System \spro\ pattern, normalized to Ba.
These patterns are taken from 
\citet{sneden08} for all elements except 
Pb and Bi, which are taken from the model predictions
presented by \citet{bisterzo11}.
The bottom panels show the abundance residuals 
with respect to the \rpro\ pattern.
}
\end{figure*}

Figure~\ref{logeplot1}
combines our results with those of
\citet{roederer12d} to 
illustrate the heavy-element abundance 
patterns of \hda\ and \hdb.
The Solar System \rpro\ and \spro\ patterns
\citep{sneden08}
are shown as fiducials for comparison.
The bottom panels of Figure~\ref{logeplot1}
illustrate the differences
between the derived abundances in each star
and the solar \rpro\ residuals
when normalized to Eu.
A relatively smooth 
pattern emerges among the differences in \hda.
The Ge, As, and Se
(32~$\leq Z \leq$~34) abundances in \hda\
exhibit an upward trend 
relative to the scaled Solar System \rpro\ component.
This trend flattens for Sr through Mo
(38~$\leq Z \leq$~42), and
the differences then smoothly decrease
for Ru through Cd
(44~$\leq Z \leq$~48).
There may be a gentle increase from
Cd through Ba
(48~$\leq Z \leq$~56),
indicated by Te,
and possibly extending through Pr
($Z =$~59).
Otherwise, the abundances of Ba through Lu
(56~$\leq Z \leq$~71)
closely follow the scaled Solar System \rpro\ pattern.
This close correspondence may extend to 
Os and Pt
($Z =$~76 and 78)
at the \third\ \rpro\ peak.
There are several anomalous elements that do not
follow this general pattern
(Y, $Z =$~39; 
Hf, $Z =$~72;
Au, $Z =$~79),
but overall there is a
relatively smooth pattern among the differences.

Similar trends are observed in \hdb,
as shown in Figure~\ref{logeplot1},
but there is more element-to-element scatter
obscuring them.
The upper limits on Os and Pt
at the \third\ \rpro\ peak
indicate a downturn in the abundances in \hdb\
relative to the scaled Solar System \rpro\ pattern.

\subsection{The 2$^{\rm nd}$ $r$-process peak}
\label{secondpeak}

There is ample observational evidence 
for the
existence of at least two components of
\rpro\ nucleosynthesis
(the so-called main and weak components)
operating in the early Galaxy
(e.g., \citealt{mcwilliam98,johnson02,aoki05,qian08,hansen12}).
Elements through Cd
($A \leq$~116) are commonly associated with
the weak component, and
the consistent \rpro\ abundance pattern found for Ba
and the rare-earth elements
($A \geq$~135) is a defining characteristic of 
the main component.

Te (125~$\leq A \leq$~130), an element at the
\second\ \rpro\ peak, 
has been detected in four stars:
\bd, 
\hda,
\hdb, and
\hdc\
(\citealt{roederer12b,roederer12a,roederer12d};
this study).
The dispersions among the ratios of 
Te and lighter elements
([Te/Ru], [Te/Pd], [Te/Ag], and [Te/Cd])
are 0.14, 0.19, 0.08, and 0.28~dex, respectively.
The dispersions among the ratios of 
Te and heavier elements
([Te/Ba], [Te/La], [Te/Ce], and [Te/Nd])
are 0.31, 0.16, 0.12, and 0.13~dex, respectively.
There is no obvious preference for smaller dispersions
among Te and lighter or heavier element groups,
which might be expected if Te
was predominantly produced in one of the 
weak or main components.
We interpret these data to indicate
that Te
owes its nucleosynthesis to conditions intermediate
between the main and weak components.

\subsection{The 1$^{\rm st}$ $r$-process peak}
\label{firstpeak}

Figure~\ref{logeplot1} indicates that the
Se abundances in \hda\ and \hdb\ are a good match to the
Solar System \rpro\ residuals when normalized to Eu.
We can estimate the percentage of Se that
originated in the main and weak components of the \rpro.
We need to consider only the differential abundance ratios
in \hda\ and \hdb\
because the stellar parameters
of these two stars are so similar.

We subtract the main component of the \rpro\ pattern
(given by the star \cs; \citealt{sneden03,sneden09}),
normalized to Eu,
from both \hda\ and \hdb.
The residual abundance pattern is attributed to the 
weak component of the \rpro\ (cf.\ \citealt{montes07}).
The weak-to-main ratio in \hda\ is 
3.7~$\pm$~0.5
times greater than in \hdb\
for elements between the \first\ and \second\ 
\rpro\ peaks. 
The ratio of the Se/Eu abundance ratios in \hda\ and \hdb\
would be low (1.0)
if the Se originated 
only in the main component of \rpro, and it would be 
high (3.7)
if the Se originated only in the weak component.
The observed ratio of the Se/Eu ratios is 
$\approx$1.4$^{+1.4}_{-0.7}$.
This is compatible with 1.0 but 
is mildly incompatible with 3.7~$\pm$~0.5,
hinting that a portion of the
Se could have originated in the main \rpro.
The 1~$\sigma$ upper limit of 2.8 
on the ratio of Se/Eu ratios
translates into 
a lower limit of the contribution from the 
main \rpro\ to Se of about 12~$\pm$~4\%.

Tighter observational uncertainties are the key to
constraining this value better
and disentangling the 
weak and main \rpro\ contributions
to the elements in the
$A \approx$~80 mass region. 
Great effort by experimentalists
has provided much of the necessary nuclear data
(masses, decay half-lives, and $\beta$-delayed neutron emission)
for each of the isotopes in this mass region
relevant for \rpro\ nucleosynthesis
(e.g., \citealt{baruah08}; \citealt{hosmer10}).
Observers must provide a suitable set of observational 
constraints for \rpro\ models that make use of these data.
The Solar System \rpro\ residuals
implicitly include contributions from all processes
except for the \spro, 
yet \citet{farouqi10} and others have pointed out that
the \rpro\ residuals in the 
$A \lesssim$~90 mass regime
are challenging to interpret because 
several nucleosynthetic processes contribute to them.
Alternate template \rpro\ patterns
for this mass regime are desirable.
The abundance pattern for elements with 32~$\leq Z \leq$~48 
in \hda\ may present an alternative.
The low metallicity of this star 
indicates that its metals derive from
relatively few enrichment events.
This star has received no or minimal \spro\ contributions.
The detection of Se, an element at the \first\
\rpro\ peak, in this star is an essential step
in the right direction.

\subsection{The beginning of nucleosynthesis beyond the iron group}

\begin{figure}
\begin{center}
\includegraphics[angle=0,width=3.2in]{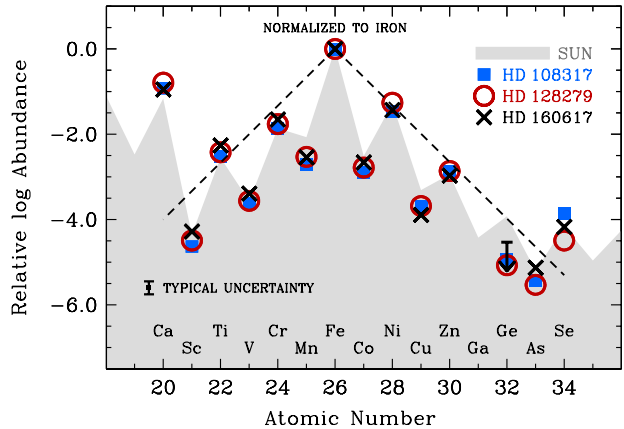}
\end{center}
\caption{
\label{fepeak}
Abundances at the iron peak in the Sun and 
three metal-poor stars.
All abundance patterns are normalized to 
$\log$~(Fe)~$=$~0.
The dashed lines are intended to guide the eye
to approximately connect the abundances of even-$Z$ elements.
Note that the Se abundances in these stars
are enhanced relative to a simple extrapolation 
of the even-$Z$ iron-peak abundances.
}
\end{figure}

The Ge-As-Se trio of elements has been studied 
in three metal-poor stars:\ \hda, \hdb, and \hdc.
Abundances of elements in the iron group
have also been reported for these three stars
by \citet{roederer12b} and \citet{roederer14}.
These abundances are illustrated in Figure~\ref{fepeak}.
The solar abundances are shown for comparison.
Each pattern is
normalized to Fe.
The dashed lines in Figure~\ref{fepeak}
trace the abundances of the even-numbered
elements in the iron group.

In massive stars,
the iron-group elements are formed during explosive $^{28}$Si burning,
and this produces the
familiar shape of the iron peak.
At the light end of the iron peak, 
Ca in the Sun and these three stars
is $>$~3~dex more abundant than
the simple extrapolation of the even-$Z$ iron peak
abundances would predict for $Z =$~20.
This results from a different,
more efficient mechanism of producing $^{40}$Ca,
$\alpha$-capture on the products of $^{16}$O burning.
At the heavy end of the iron peak, 
an extrapolation of the dashed line to elements heavier than Zn
reveals that the stellar Ge abundances fall
$\approx$~1~dex lower than the extrapolation,
and the stellar Se abundances fall
$\approx$~1~dex higher than the extrapolation
These values disagree with the extrapolation by 
$\approx$~2.5--3.5~times the abundance uncertainties.
Using the Ca overabundance to guide our 
interpretation of Ge and Se,
we speculate that the 
stellar deficiency of Ge and excess of Se
indicate the mass regime 
(75~$< A \leq$~82) where
a different nucleosynthetic mechanism
dominates the quasi-equilibrium 
$\alpha$-rich freezeout of the iron peak.

In general,
\rpro\ models predict that these elements are produced by 
charged-particle reactions,
neutron-capture reactions, or both.
Both $\nu p$ reactions and $\alpha$-rich freezeout
produce
low-$A$, neutron-deficient isotopes
\citep{frohlich06,pruet06}, while 
neutron-capture reactions
produce high-$A$, neutron-rich isotopes
\citep{woosley92,woosley94,farouqi10}.
The fact that the excess occurs at Se ($A \approx$~80)
may provide a clue to the dominant nucleosynthesis mechanism.
Nuclei in this mass region could 
be the stable descendants
of unstable nuclei at the 
neutron-magic $N =$~50 closed shell along the \rpro\ path,
hinting at an \rpro\ origin.
Isotopic abundances of Ge and Se would 
resolve the issue definitively, but
measurements of isotopic ratios are
not likely to be available in the foreseeable future.
Ge and Se have small hfs and IS,
making such measurements difficult if not impossible,
and in any case the NUV spectroscopic observations
that would be required are prohibitively long.
Two practical approaches to better understand
the astrophysical origins of these nuclei
may be to derive elemental abundances from
larger samples of
stars with diverse heavy-element abundance patterns
and to reduce the statistical uncertainties
by acquiring higher-quality spectra when possible.

\section{Summary}
\label{summary}

We have obtained new high-resolution, high-S/N NUV spectroscopic
observations of the metal-poor giants \hda\ and \hdb.
These observations extend the high-quality 
spectral coverage of \hda\ and \hdb\
from $\lambda <$~2000~\AA\ to nearly 1~$\mu$m.
We have derived new abundances or upper limits
from 27~lines of elements with $Z \geq$~29 in these stars.

We have derived [Cu/Fe] from both neutral and ionized species, and
$\langle$[Cu~\textsc{ii}/Cu~\textsc{i}]$\rangle = +$0.56~$\pm$~0.23
in \hda\ and \hdb.
These data hint that
Cu~\textsc{i} may not be formed in LTE in these stars.
We are not aware of any non-LTE calculations
for Cu in late-type stars.
We echo the caution issued by \citet{bonifacio10} that
Cu abundances derived from Cu~\textsc{i} lines
in metal-poor stars should be treated with caution
until such calculations become available.

The [Zn/Fe], [Mo/Fe], [Cd/Fe], and [Os/Fe]
ratios have been derived from both neutral and ionized species.
In these cases the ratios from neutrals and ions
agree within the uncertainties, which span
0.15 to 0.52~dex
(see Table~\ref{iontab}).
These data also
corroborate the Ge, Cd, Te, Os, and Pt abundances
derived from other 
lines in \hda\ and \hdb\
\citep{roederer12a,roederer12d}.

Comparison of the As and Se abundances with those of iron-group elements
indicates that
the mass region between Ge and Se 
(75~$\leq A \leq$~82) may identify the
mass at which a different nucleosynthetic mechanism
begins to dominate the quasi-equilibrium
$\alpha$-rich freezeout of the iron peak.

The data hint that a portion ($\gtrsim$~10\%) of the
Se could have originated in the main component of the \rpro.
Reducing both the statistical and systematic uncertainties
on Se and
increasing the sample of stars with precise As and Se
measurements should remain high priorities
while \textit{HST} is operational.

\acknowledgments

We thank the referee for providing helpful comments on this manuscript.
This research has made use of NASA's 
Astrophysics Data System Bibliographic Services, 
the arXiv pre-print server operated by Cornell University, 
the SIMBAD and VizieR databases hosted by the
Strasbourg Astronomical Data Center,
the ASD hosted by NIST, and
the Mikulski Archive at the Space Telescope Science Institute.
IRAF is distributed by the National Optical Astronomy Observatories,
which are operated by the Association of Universities for Research
in Astronomy, Inc., under cooperative agreement with the National
Science Foundation.
Generous support for Programs GO-12268 and GO-12976 was
provided by NASA through
grants from the Space Telescope Science Institute, which is operated by the
Association of Universities for Research in Astronomy, Incorporated, under
NASA contract NAS~5-26555.
HS acknowledges support from grant
PHY~11-0251 from the US National Science Foundation.
HS and TCB acknowledge partial support for this work 
from grant PHY~08-22648; 
Physics Frontier Center / 
Joint Institute for Nuclear Astrophysics (JINA), 
awarded by the US National Science Foundation.
AF acknowledges support from grant AST~12-55160
from the US National Science Foundation.

 {\it Facilities:} 
\facility{HST (STIS)}

\end{document}

%% file: tab1.tex
\begin{deluxetable*}{cccccccccc}
\tablecaption{Iron Equivalent Widths and Abundances
\label{ewtab}}
\tablewidth{0pt}
\tabletypesize{\scriptsize}
\tablehead{
\colhead{} &
\colhead{} &
\colhead{} &
\colhead{} &
\colhead{} &
\multicolumn{2}{c}{HD 108317} &
\colhead{} &
\multicolumn{2}{c}{HD 128279} \\
\cline{6-7} \cline{9-10}
\colhead{Species} &
\colhead{$\lambda$} &
\colhead{E.P.} &
\colhead{\loggf} &
\colhead{NIST} &
\colhead{Eq.\ Wid.} &
\colhead{$\log \epsilon$} &
\colhead{} &
\colhead{Eq.\ Wid.} &
\colhead{$\log \epsilon$} \\
\colhead{} &
\colhead{(\AA)} &
\colhead{(eV)} &
\colhead{} &
\colhead{Grade\tablenotemark{a}} &
\colhead{(m\AA)} &
\colhead{} &
\colhead{} &
\colhead{(m\AA)} &
\colhead{} }
\startdata
 Fe~\textsc{i}  &  2132.02 &  0.00 & $-$1.33 & C+ & \nodata &\nodata&  &    90.1 &  4.44 \\ 
 Fe~\textsc{i}  &  2145.19 &  0.05 & $-$1.56 & C  &    79.8 &  4.56 &  &    82.2 &  4.56 \\ 
 Fe~\textsc{i}  &  2161.58 &  0.11 & $-$1.76 & D+ &    83.0 &  4.89 &  &    86.1 &  4.89 \\ 
 Fe~\textsc{i}  &  2176.84 &  0.12 & $-$1.66 & C+ & \nodata &\nodata&  &    91.4 &  4.90 \\ 
 Fe~\textsc{i}  &  2191.20 &  0.12 & $-$1.80 & C  &    72.0 &  4.65 &  &    70.0 &  4.51 \\ 
 Fe~\textsc{i}  &  2196.04 &  0.11 & $-$0.59 & C+ &   144.2 &  4.45 &  &   156.3 &  4.51 \\ 
 Fe~\textsc{i}  &  2200.72 &  0.11 & $-$0.99 & C  & \nodata &\nodata&  &   140.2 &  4.79 \\ 
 Fe~\textsc{i}  &  2228.17 &  0.05 & $-$2.11 & D+ &    81.4 &  5.11 &  &    79.5 &  5.00 \\ 
 Fe~\textsc{i}  &  2259.28 &  0.05 & $-$2.31 & C  &    60.2 &  4.62 &  & \nodata &\nodata\\ 
 Fe~\textsc{i}  &  2259.51 &  0.00 & $-$1.32 & B+ & \nodata &\nodata&  &   187.4 &  5.32 \\ 
 Fe~\textsc{i}  &  2265.05 &  0.05 & $-$2.11 & C  &    96.3 &  5.38 &  &    91.0 &  5.24 \\ 
 Fe~\textsc{i}  &  2267.08 &  0.05 & $-$1.75 & B  & \nodata &\nodata&  &    91.8 &  4.88 \\ 
 Fe~\textsc{i}  &  2275.19 &  0.11 & $-$2.32 & D+ &    71.3 &  5.07 &  &    69.3 &  4.93 \\ 
 Fe~\textsc{i}  &  2283.30 &  0.12 & $-$2.22 & B+ &    92.7 &  5.50 &  & \nodata &\nodata\\ 
 Fe~\textsc{i}  &  2294.41 &  0.11 & $-$1.54 & B  &    83.6 &  4.63 &  &    86.0 &  4.62 \\ 
 Fe~\textsc{i}  &  2297.79 &  0.05 & $-$1.10 & D+ &    96.7 &  4.37 &  &   113.3 &  4.56 \\ 
 Fe~\textsc{i}  &  2298.66 &  0.11 & $-$2.42 & C  &    82.5 &  5.48 &  & \nodata &\nodata\\ 
 Fe~\textsc{i}  &  2299.22 &  0.09 & $-$1.55 & C  &    76.4 &  4.43 &  &    85.2 &  4.57 \\ 
 Fe~\textsc{i}  &  2309.00 &  0.11 & $-$1.39 & C  &   106.0 &  4.97 &  &   118.7 &  5.09 \\ 
 Fe~\textsc{ii} &  2122.45 &  1.96 & $-$2.59 & D+ &    41.2 &  5.06 &  &    42.4 &  5.00 \\ 
 Fe~\textsc{ii} &  2162.02 &  1.96 & $-$0.75 & C  &    87.2 &  4.90 &  &    88.7 &  4.85 \\ 
 Fe~\textsc{ii} &  2209.03 &  4.77 & $-$0.13 & C+ & \nodata &\nodata&  &    28.6 &  4.98 \\ 
 Fe~\textsc{ii} &  2251.55 &  0.05 & $-$2.35 & D  &   156.8 &  5.03 &  &   159.9 &  5.00 \\ 
 Fe~\textsc{ii} &  2254.41 &  0.11 & $-$3.07 & D  &    78.0 &  4.83 &  & \nodata &\nodata\\ 
 Fe~\textsc{ii} &  2260.86 &  0.11 & $-$2.00 & C+ &   145.4 &  4.84 &  &   142.6 &  4.75 \\ 
 Fe~\textsc{ii} &  2262.69 &  0.11 & $-$2.22 & C+ &   126.0 &  4.71 &  &   134.5 &  4.72 \\ 
 Fe~\textsc{ii} &  2268.56 &  0.12 & $-$2.73 & C  &   104.5 &  5.16 &  &   103.1 &  5.05 \\ 
 Fe~\textsc{ii} &  2268.82 &  0.05 & $-$2.61 & C+ &   114.7 &  5.02 &  &   118.2 &  4.99 \\ 
 Fe~\textsc{ii} &  2351.20 &  2.66 & $-$0.23 & C+ &    97.2 &  5.23 &  &    96.0 &  5.14 \\ 
 Fe~\textsc{ii} &  2351.67 &  5.25 & $-$0.05 & C  & \nodata &\nodata&  &    21.3 &  5.14 \\ 
\enddata
\tablenotetext{a}{
NIST grades correspond to accuracies of 
$\leq$~7\% (B$+$), 
10\% (B), 
18\% (C$+$),
25\% (C),
40\% (D$+$), and
50\% (D).
}
\end{deluxetable*}

%% file: tab2.tex
\begin{deluxetable}{ccccc}
\tablecaption{Mean Fe~\textsc{i} Abundances Binned by Wavelength
\label{corrtab}}
\tablewidth{0pt}
\tabletypesize{\scriptsize}
\tablehead{
\colhead{Wavelength} &
\colhead{$\langle\log\epsilon\rangle$} &
\colhead{Std.\ Dev.} &
\colhead{N} &
\colhead{Correction} \\
\colhead{Range (\AA)} &
\colhead{} &
\colhead{} &
\colhead{} &
\colhead{} }
\startdata
\multicolumn{5}{c}{HD~108317} \\
\hline
$<$ 2360   & 4.87 & 0.39 & 9   & $+$0.10     \\
2360--3100 & 4.93 & 0.16 & 32  & $+$0.04     \\
3100--3647 & 4.81 & 0.09 & 107 & $+$0.16     \\
3647--4000 & 4.87 & 0.08 & 40  & $+$0.10     \\
4000--4400 & 4.92 & 0.07 & 27  & $+$0.05     \\
4400--6750 & 4.97 & 0.07 & 94  & $\equiv$0.0 \\
\hline\hline
\multicolumn{5}{c}{HD~128279} \\
\hline
$<$ 2360   & 4.80 & 0.27 & 10 & $+$0.22     \\
2360--3100 & 4.91 & 0.17 & 28 & $+$0.11     \\
3100--3647 & 4.83 & 0.10 & 96 & $+$0.19     \\
3647--4000 & 4.91 & 0.07 & 32 & $+$0.11     \\
4000--4400 & 4.97 & 0.07 & 19 & $+$0.05     \\
4400--6750 & 5.02 & 0.07 & 91 & $\equiv$0.0 \\
\enddata
\end{deluxetable}

%% file: tab3.tex
\begin{deluxetable*}{cccccccc}
\tablecaption{Atomic Data and Line Abundances
\label{atomictab}}
\tablewidth{0pt}
\tabletypesize{\scriptsize}
\tablehead{
\colhead{Species} &
\colhead{$Z$} &
\colhead{Wavelength\tablenotemark{a}} &
\colhead{E.P.} &
\colhead{\loggf\ ($\sigma$)} &
\colhead{Ref.} &
\colhead{$\log \epsilon$ ($\sigma$)} &
\colhead{$\log \epsilon$ ($\sigma$)} \\
\colhead{} &
\colhead{} &
\colhead{(\AA)} &
\colhead{(eV)} &
\colhead{} &
\colhead{} &
\colhead{HD 108317} &
\colhead{HD 128279} }
\startdata
Cu~\textsc{ii} & 29 & 2037.13 & 2.83 & $-$0.23 (0.03) & 1  & $+$1.48 (0.15) & $+$1.21 (0.30) \\  
Cu~\textsc{ii} & 29 & 2054.98 & 2.93 & $-$0.29 (0.03) & 1  & $+$1.57 (0.20) & $+$1.45 (0.30) \\  
Cu~\textsc{ii} & 29 & 2112.10 & 3.25 & $-$0.11 (0.10) & 1  & $+$1.24 (0.20) & $+$1.42 (0.30) \\  
Cu~\textsc{ii} & 29 & 2126.04 & 2.83 & $-$0.23 (0.07) & 1  & $+$1.39 (0.20) & \nodata        \\
Zn~\textsc{ii} & 30 & 2062.00 & 0.00 & $-$0.29 (0.04) & 2  & $+$2.50 (0.30) & $+$2.38 (0.30) \\  
Ge~\textsc{i}  & 32 & 2041.71 & 0.00 & $-$0.7  (0.10) & 3  & $+$0.53 (0.30) & \nodata        \\
Ge~\textsc{i}  & 32 & 2065.21 & 0.07 & $-$0.79 (0.10) & 3  & $<+$0.85       & \nodata        \\
As~\textsc{i}  & 33 & 1972.62 & 0.00 & $-$0.63 (0.05) & 1  & $-$0.30 (0.40) & $-$0.49 (0.40) \\  
Se~\textsc{i}  & 34 & 1960.89 & 0.00 & $-$0.43 (0.08) & 4  & $+$1.30 (0.40) & $+$0.57 (0.40) \\  
Se~\textsc{i}  & 34 & 2074.78 & 0.00 & $-$2.26 (0.03) & 4  & $+$1.25 (0.40) & $+$0.53 (0.40) \\  
Mo~\textsc{ii} & 42 & 2015.11 & 0.00 & $-$0.49 (0.11) & 5  & \nodata        & $-$0.74 (0.30) \\  
Mo~\textsc{ii} & 42 & 2020.31 & 0.00 & $+$0.02 (0.10) & 5  & $-$0.16 (0.30) & $-$0.51 (0.20) \\  
Mo~\textsc{ii} & 42 & 2045.97 & 0.00 & $-$0.35 (0.12) & 5  & $+$0.18 (0.20) & $-$0.48 (0.20) \\  
Mo~\textsc{ii} & 42 & 2660.58 & 1.49 & $-$0.14 (0.15) & 5  & $-$0.15 (0.20) & \nodata        \\
Cd~\textsc{i}  & 48 & 2288.02 & 0.00 & $+$0.15 (0.02) & 4  & $-$1.21 (0.20) & $-$1.34 (0.30) \\  
Cd~\textsc{ii} & 48 & 2144.39 & 0.00 & $+$0.02 (0.02) & 1  & $-$1.08 (0.30) & $-$1.30 (0.30) \\  
Te~\textsc{i}  & 52 & 2142.82 & 0.00 & $-$0.32 (0.08) & 6  & $+$0.14 (0.30) & $-$0.04 (0.30) \\  
Yb~\textsc{ii} & 70 & 2116.68 & 0.00 & $-$1.34 (0.10) & 1  & $<-$0.50       & $<-$0.98       \\  
W~\textsc{ii}  & 74 & 2088.20 & 0.39 & $-$0.02 (0.03) & 7  & $<-$0.80       & $<-$0.58       \\
W~\textsc{ii}  & 74 & 2118.88 & 0.00 & $-$0.77 (0.05) & 7  & $<-$0.40       & $<-$0.68       \\
Os~\textsc{ii} & 76 & 2067.23 & 0.45 & $-$0.05 (0.03) & 8  & $-$0.47 (0.30) & $<-$0.38       \\
Os~\textsc{ii} & 76 & 2282.28 & 0.00 & $-$0.05 (0.04) & 9  & $-$0.83 (0.15) & $<-$1.28       \\
Pt~\textsc{i}  & 78 & 2049.39 & 0.00 & $+$0.02 (0.06) & 1  & $-$0.42 (0.20) & $<-$0.53       \\
Pt~\textsc{i}  & 78 & 2067.51 & 0.00 & $-$0.62 (0.03) & 10 & $+$0.01 (0.30) & $<+$0.02       \\
Pt~\textsc{i}  & 78 & 2144.21 & 0.00 & $-$0.37 (0.07) & 11 & $-$0.37 (0.20) & $<-$0.78       \\
Hg~\textsc{ii} & 80 & 1942.27\tablenotemark{b} 
                              & 0.00 & $-$0.40 (0.04) & 1  & $<+$0.30       & $<+$0.53       \\
Pb~\textsc{ii} & 82 & 2203.53 & 1.75 & $-$0.14 (0.10) & 3  & $<+$0.09       & $<+$0.12       \\
Bi~\textsc{i}  & 83 & 2230.61 & 0.00 & $+$0.02 (0.06) & 12 & $<-$1.02       & \nodata        \\
\enddata
\tablenotetext{a}{
Air wavelengths are given for $\lambda >$~2000~\AA\ and vacuum values below.
}
\tablenotetext{b}{
The wavelength of this line was incorrectly listed 
in Table~14 of \citet{roederer12b}.  
The value given here is correct.
}
\tablerefs{
(1) \citealt{roederer12b};  
(2) \citealt{bergeson93};
(3) NIST; 
(4) \citealt{morton00};
(5) \citealt{sikstrom01};
(6) \citealt{roederer12a};  
(7) \citealt{nilsson08};
(8) \citealt{ivarsson04};
(9) \citealt{quinet06};
(10) \citealt{denhartog05};
(11) This study;
(12) \citealt{morton00}
}
\end{deluxetable*}

%% file: tab4.tex
\begin{deluxetable*}{cccccccccccccc}
\tablecaption{Mean Abundances of Species Examined in This Study
\label{abundtab}}
\tablewidth{0pt}
\tabletypesize{\scriptsize}
\tablehead{
\colhead{} &
\colhead{} &
\colhead{} &
\multicolumn{5}{c}{HD 108317} &
\colhead{} &
\multicolumn{5}{c}{HD 128279} \\
\cline{4-8} \cline{10-14}
\colhead{Species} &
\colhead{$Z$} &
\colhead{$\log \epsilon_{\odot}$} &
\colhead{$\log \epsilon$} &
\colhead{[X/Fe]} &
\colhead{$\sigma_{\rm stat}$} &
\colhead{$\sigma_{\rm total}$} &
\colhead{N} &
\colhead{} &
\colhead{$\log \epsilon$} &
\colhead{[X/Fe]} &
\colhead{$\sigma_{\rm stat}$} &
\colhead{$\sigma_{\rm total}$} &
\colhead{N}
}
\startdata
Cu~\textsc{ii} & 29 & 4.19 &  $+$1.42 &  $-$0.40 &   0.22  &   0.28  & 4                  & &  $+$1.36 &  $-$0.37 &   0.23  &   0.29  & 3                  \\
Zn~\textsc{ii} & 30 & 4.56 &  $+$2.50 &  $+$0.31 &   0.49  &   0.52  & 1                  & &  $+$2.38 &  $+$0.28 &   0.41  &   0.44  & 1                  \\
Ge~\textsc{i}  & 32 & 3.65 &  $+$0.19 &  $-$0.93 &   0.14  &   0.18  & 4\tablenotemark{a} & &  $-$0.03 &  $-$1.20 &   0.25  &   0.28  & 3\tablenotemark{a} \\
As~\textsc{i}  & 33 & 2.30 &  $-$0.30 &  $-$0.07 &   0.56  &   0.57  & 1                  & &  $-$0.49 &  $-$0.31 &   0.49  &   0.50  & 1                  \\
Se~\textsc{i}  & 34 & 3.34 &  $+$1.28 &  $+$0.47 &   0.40  &   0.42  & 2                  & &  $+$0.55 &  $-$0.31 &   0.34  &   0.36  & 2                  \\
Mo~\textsc{ii} & 42 & 1.88 &  $-$0.05 &  $+$0.44 &   0.28  &   0.33  & 3                  & &  $-$0.56 &  $+$0.02 &   0.22  &   0.28  & 3                  \\
Cd~\textsc{i}  & 48 & 1.71 &  $-$1.21 &  $-$0.39 &   0.44  &   0.46  & 1                  & &  $-$1.34 &  $-$0.57 &   0.34  &   0.36  & 1                  \\
Cd~\textsc{ii} & 48 & 1.71 &  $-$1.08 &  $-$0.42 &   0.49  &   0.52  & 1                  & &  $-$1.30 &  $-$0.55 &   0.40  &   0.43  & 1                  \\
Te~\textsc{i}  & 52 & 2.18 &  $+$0.06 &  $+$0.41 &   0.28  &   0.30  & 2\tablenotemark{a} & &  $-$0.12 &  $+$0.18 &   0.27  &   0.30  & 2\tablenotemark{a} \\
Yb~\textsc{ii} & 70 & 0.92 &  $-$1.13 &  $+$0.32 &   0.14  &   0.22  & 1\tablenotemark{a} & &  $-$1.69 &  $-$0.15 &   0.13  &   0.21  & 1\tablenotemark{a} \\
W~\textsc{ii}  & 74 & 0.65 & $<-$0.80 & $<+$0.92 & \nodata & \nodata & 2                  & & $<-$0.68 & $<+$1.13 & \nodata & \nodata & 2                  \\
Os~\textsc{ii} & 76 & 1.40 &  $-$0.68 &  $+$0.29 &   0.32  &   0.36  & 2                  & & $<-$1.28 & $<-$0.22 & \nodata & \nodata & 2                  \\
Pt~\textsc{i}  & 78 & 1.62 &  $-$0.40 &  $+$0.51 &   0.15  &   0.19  & 5\tablenotemark{a} & & $<-$0.78 & $<+$0.08 & \nodata & \nodata & 5\tablenotemark{a} \\
Hg~\textsc{ii} & 80 & 1.17 & $<+$0.30 & $<+$1.50 & \nodata & \nodata & 1                  & & $<+$0.53 & $<+$1.82 & \nodata & \nodata & 1                  \\
Pb~\textsc{ii} & 82 & 2.04 & $<+$0.09 & $<+$0.42 & \nodata & \nodata & 1                  & & $<+$0.12 & $<+$0.54 & \nodata & \nodata & 1                  \\
Bi~\textsc{i}  & 83 & 0.65 & $<-$1.02 & $<+$0.86 & \nodata & \nodata & 1                  & & \nodata  & \nodata  & \nodata & \nodata & \nodata            \\
\enddata
\tablecomments{
The [X/Fe] ratios, where
X represents one of the heavy elements of interest,
are calculated by comparing 
neutrals with neutrals and ions with ions.
}
\tablenotetext{a}{
The mean abundance or upper limit includes one or more lines 
from \citet{roederer12d}
}
\end{deluxetable*}

%% file: tab5.tex
\begin{deluxetable}{ccc}
\tablecaption{Comparison of [X/Fe] Ratios of Heavy Elements
Derived from Neutral and Ionized Species
\label{iontab}}
\tablewidth{0pt}
\tabletypesize{\scriptsize}
\tablehead{
\colhead{Species} &
\colhead{HD 108317} &
\colhead{HD 128279} 
}
\startdata
Cu~\textsc{i}  & $-$0.96~$\pm$~0.16 & $-$0.92~$\pm$~0.16 \\
Cu~\textsc{ii} & $-$0.40~$\pm$~0.28 & $-$0.37~$\pm$~0.29 \\
\hline
Zn~\textsc{i}  & $+$0.23~$\pm$~0.15 & $+$0.10~$\pm$~0.15 \\
Zn~\textsc{ii} & $+$0.31~$\pm$~0.52 & $+$0.28~$\pm$~0.44 \\
\hline
Mo~\textsc{i}  & $+$0.42~$\pm$~0.19 & $-$0.07~$\pm$~0.18 \\
Mo~\textsc{ii} & $+$0.44~$\pm$~0.33 & $+$0.02~$\pm$~0.28 \\
\hline
Cd~\textsc{i}  & $-$0.39~$\pm$~0.46 & $-$0.57~$\pm$~0.36 \\
Cd~\textsc{ii} & $-$0.42~$\pm$~0.52 & $-$0.55~$\pm$~0.43 \\
\hline
Os~\textsc{i}  & $+$0.63~$\pm$~0.23 & $<+$0.69           \\
Os~\textsc{ii} & $+$0.29~$\pm$~0.36 & $<-$0.22           \\
\hline
Pb~\textsc{i}  & $<+$0.23           & $<+$0.35           \\
Pb~\textsc{ii} & $<+$0.42           & $<+$0.54           \\
\enddata
\end{deluxetable}